\documentclass{aa}  

\usepackage{graphicx}

\usepackage{txfonts}

\usepackage{amsmath}
\usepackage{multirow}
\usepackage[section]{placeins}
\usepackage{hyperref}
\usepackage{mathtools}
\usepackage{xcolor}
\usepackage{soul}
\usepackage{ulem}

\makeatletter
\renewcommand*\aa@pageof{, page \thepage{} of \pageref*{LastPage}}
\makeatother

\hypersetup{colorlinks=true,allcolors=[rgb]{0,0,0.8}}

\newcommand{\simba}{{\sc Simba}}
\newcommand{\gizmo}{{\sc Gizmo}}
\newcommand{\gadget}{{\sc GADGET-3}}
\newcommand{\caesar}{{\sc Caesar}}
\newcommand{\yt}{{\sc yt}}
\newcommand{\grackle}{{\sc Grackle-3.1}}
\newcommand{\disperse}{{\sc DisPerSE}}

\usepackage{orcidlink}
\newcommand{\orcid}[1]{\orcidlink{#1}}

\begin{document}

   \title{Star-forming galaxies in the cosmic web in the last 11 Gyr}
   \authorrunning{Jego, B., et al.}
   
   \author{Baptiste Jego\orcid{0009-0006-6399-7858}
          \inst{1},
          Katarina Kraljic\orcid{0000-0001-6180-0245}
          \inst{1},
          Matthieu Béthermin\orcid{0000-0002-3915-2015}
          \inst{1},
          \and 
          Romeel Davé\orcid{0000-0003-2842-9434}
          \inst{2, 3}
          }

   \institute{Observatoire Astronomique de Strasbourg, UMR 7550, CNRS, Université de Strasbourg, F-67000 Strasbourg, France\\
   \email{baptiste.jego@astro.unistra.fr}
   \and
   Institute for Astronomy, University of Edinburgh, Royal Observatory, Edinburgh EH9 3HJ, UK
   \and
   Department of Physics and Astronomy, University of the Western Cape, Robert Sobukwe Rd, Cape Town 7535, South Africa
            }

    \date{Received 22 September 2025 / Accepted 21 December 2025}
 
  \abstract
   {
   We investigate how the star formation activity of galaxies depends on their position within the cosmic web using the \simba~cosmological simulation from redshift $z=3$ to $z=0$. While previous studies found that galaxies closer to filaments tend to be more massive and quenched, it remained unclear whether these trends reflect intrinsic environmental effects or changes in the galaxy population mix. To address this, we focus exclusively on star-forming galaxies, robustly selected using both the specific star formation rate (sSFR) and gas depletion timescale criteria, in order to isolate the direct impact of the cosmic web on star-forming galaxies. We reconstruct the 3D cosmic web skeleton using \disperse~and compute each galaxy's distance to its nearest filament. After explicitly removing the stellar-mass dependence of all quantities, we examine deviations in star formation rate (SFR), sSFR, molecular and atomic gas depletion timescales, and gas fractions as a function of this distance. We find a clear and redshift-dependent modulation of star formation with filament proximity: at high redshift ($z \gtrsim 2$), galaxies closer to filaments show enhanced SFR and gas accretion, reflecting efficient filament-fed growth. At $z=0$, we observe a V-shaped trend in the sSFR and depletion timescales, with minima at intermediate distances ($\sim 0.25$ cMpc) and a surprising upturn very close to the filament cores, suggesting a resumed accretion in the densest environments. These effects are not driven by mergers and are primarily associated with satellite galaxies at low redshift. Our results demonstrate that large-scale cosmic web proximity modulates star formation in star-forming galaxies through a combination of gas supply regulation and environmental processing, with different mechanisms dominating across cosmic time.
   }


   \maketitle

\section{Introduction}\label{sec:intro}

The understanding of galaxy formation and evolution is still facing major challenges, among which is the role of the large-scale environment in shaping various galaxy properties. The mass assembly of galaxies follows the dark matter halos assembly history at the first order \citep[e.g.][]{1984Natur.311..517B, 1996ApJ...462..563N}, but is also intricately linked to the interplay between internal processes and the cosmic environment in which galaxies reside. This larger-scale environment, known as the cosmic web \citep{1996Natur.380..603B}, is composed of voids, walls, filaments, and nodes where galaxies form, grow, and interact. Its emergence from the gravitational amplification of initial density perturbations can be intuitively understood through the Zel'dovich approximation \citep{1970A&A.....5...84Z}.

On the observational side, early signs of coherent large-scale structures appeared in the late 1970s - early 1980s (e.g. \citealt{gregory1978coma, 10.1093/mnras/185.2.357,1980MNRAS.193..353E, 1981ApJ...248L..57K}), then galaxy redshift surveys have mapped the spatial distribution of galaxies and revealed the web of filaments and nodes up to redshift $z\sim0.9$ (e.g. \citealt{1986ApJ...302L...1D, York_2000, 2001MNRAS.328.1039C, 2004PhRvD..69j3501T, 10.1111/j.1365-2966.2009.15470.x, 2011MNRAS.413..971D, 2014A&A...566A.108G, malavasi2017, 10.1093/mnras/sty1553}), and future surveys will allow us to extend these efforts up to redshift $z\sim2$ (e.g. \citealt{2014PASJ...66R...1T, euclid}). Beyond redshift $z>2$, the large-scale structure is primarily accessible through tomographic reconstruction using the absorption features (e.g. Lyman-$\alpha$ forest) in the spectra of bright background sources such as quasars and star-forming galaxies (e.g. \citealt{Lee_2016}). However, modern hydrodynamical and N-body simulations such as Horizon-AGN \citep{2014MNRAS.444.1453D}, IllustrisTNG \citep{2018MNRAS.473.4077P}, or \simba~\citep{Dav__2019} have further quantified the roles these structures play in galaxy evolution, providing insight into the interplay between baryonic and dark-matter dynamics.

Galaxies located near dense cosmic structures, such as filaments and nodes, exhibit distinct properties compared to those in less dense environments, possibly reflecting the role of gravitational dynamics, gas accretion, and feedback. Matter is expelled from voids towards walls and filaments where it flows towards nodes, leading to anisotropic accretion that shapes both halo and galaxy properties. This results in higher stellar masses, quenched star formation, and altered gas and metallicity distributions near filaments, influenced by tidal forces and pre-processing \citep[e.g.][]{10.1093/mnras/stw3127, kuutma2017, malavasi2017, kraljic2018, laigle2018, Hasan_2023, 10.1093/mnras/stae667}. Filaments also align galaxy shapes and spins through coherent tidal fields \citep[e.g.][]{10.1111/j.1365-2966.2009.15271.x, Codis2015, Krolewski2019, 2020MNRAS.493..362K}. Additionally, the geometry and dynamics of filaments modulate accretion and halo shapes, contributing to assembly bias \citep[e.g.][]{10.1111/j.1365-2966.2004.07733.x, 2005MNRAS.363L..66G}.

Recent advances in cosmological simulations have enabled detailed investigations into how galaxy evolution is influenced by the cosmic web. State-of-the-art cosmological simulations, such as \simba, enable detailed investigations of galaxy properties, including stellar mass, star formation rates, gas fractions, and metallicity, while accounting for the effects of large-scale structures such as filaments and nodes. These simulations make it possible to study the dependence of galactic properties as a function of their proximity to cosmic web features, shedding light on how anisotropic gas accretion, tidal interactions, and pre-processing within filaments influence galaxy assembly and quenching. Building on this, \citet{10.1093/mnras/stae667} conducted a comprehensive analysis of how global galaxy properties, stellar mass, star formation rate (SFR), atomic and molecular gas fractions, and metallicity vary with distance to cosmic filaments as identified by \disperse~\citep{Sousbie2011, SousbiePK2011}. Their study found that galaxies located closer to filaments tend to be more massive and more quenched, exhibiting lower specific SFRs and reduced gas fractions, particularly at $z \lesssim 1$. These trends are consistent with previous observational and theoretical studies \citep[e.g.][]{Sarron2019, Bennett2020}, and have been interpreted as signatures of pre-processing within the filamentary environment, potentially driven by mechanisms such as shock heating, tidal stripping, or active galactic nuclei (AGN) feedback \citep[e.g.][]{Goldsmith2016, Irodotou2022}. Importantly, the increasing fraction of quenched galaxies near filaments also points to a shift in population mix across environments. This shift raises the question: to what extent do the environmental trends observed in galaxy populations arise from differences in galaxy type (i.e. star-forming vs. quenched), versus direct modulation of galaxy properties by the large-scale structure?

In this work, we revisit this question with a sharper focus on the star-forming phase of galaxies. We restrict our analysis to star-forming galaxies only, in order to isolate the direct influence of the cosmic web on star formation and gas content, independent of the confounding effects of quenching. While \citet{10.1093/mnras/stae667} identified global suppression of star formation near filaments, the declining fraction of star-forming galaxies in those regions makes it difficult to disentangle whether the environment directly alters the SFR of active galaxies, or simply reflects the presence of more quenched systems. By selecting only galaxies on or near the star-forming main sequence, we aim to uncover whether proximity to filaments affects SFRs, gas fractions, molecular and atomic gas depletion times, and stellar mass within the star-forming population itself. This allows us to directly test whether filaments modulate accretion and star formation efficiencies in active galaxies, and thus whether pre-processing is already at play prior to quenching. We aim to quantify the properties of star-forming galaxies within \simba~and quantify the average effect of their distance to the cosmic web. 

We use the flat $\Lambda$CDM cosmological parameters from Planck \citep{Planck2016} as implemented in the \simba~simulation: $\Omega_{m}=0.3, \Omega_{\Lambda}=0.7, \Omega_{b}=0.048, H_{0}=68$ km cMpc$^{-1}$ s$^{-1}, \sigma_{8}=0.82$ and $n_{s}=0.97$, and we refer to the logarithm in base ten as log.

The paper is structured as follows. In Sect.~\ref{sec:data} we present the data extracted from the \simba~simulation, which we model in Sect.~\ref{sec:methods}. Then, in Sect.~\ref{sec:results} we present and comment on our results, which we discuss and interpret in Sect.~\ref{sec:disc}, and we present our conclusions in Sect.~\ref{sec:conc}. Additional details on the distribution of galaxies distances to filaments, gas-mass accretion, the effect of massive halo removal, the link between the cosmic web and the extragalactic gas density, and the movement of galaxies relative to filaments are provided in Appendices~\ref{appA},~\ref{appB},~\ref{appC},~\ref{appD}, and~\ref{appE} respectively.

\section{Data}\label{sec:data}

In this section, we give a brief presentation of the data extracted from the \simba~simulation, with a focus on the most relevant quantities used in the rest of this work.

\subsection{Galaxy evolution in \simba}\label{ssec:simba}

The \simba~simulation suite is a state-of-the-art large-scale simulation of galaxy formation \citep[see][for a complete description, including numerous observables and their redshift evolution, such as galaxy stellar mass function, stellar mass-star formation rate main sequence, or gas fractions]{Dave2019}. It is a hydrodynamical simulation and includes significant updates from previous ones, especially in its treatment of black hole growth, AGN feedback, and dust production. It uses the Meshless-Finite-Mass hydrodynamic version of the \gizmo~code \citep{Hopkins2015} and the \gadget~tree-particle-mesh gravity solver \citep{Springel2005}.

\simba~contains $1024^{3}$ dark matter and gas particles, with dark matter and gas elements each with masses of $\sim 10^{7} M_{\odot}$ allowing accurate tracking of the evolution of galaxy populations from high redshift ($z=6$) to the present, capturing galaxy properties across cosmic times in a self-consistent cosmological framework. Cooling is handled with the \grackle~library \citep{Smith2017}, which calculates radiative cooling from primordial and metal-enriched gases, simulating the interaction between gas cooling and a background of ionising radiation. Self-shielding, where gas at high densities shields itself from this radiation, is approximated to ensure realistic gas densities and temperatures.

Star formation is modelled using a Schmidt-type relation \citep{Schmidt1959} based on the molecular hydrogen density, $\rho_{\text{H}_2}$, in each gas element. The SFR is given by
\begin{equation}
    \text{SFR} = \epsilon_{*} \frac{\rho_{\text{H}_2}}{t_{\text{dyn}}},
    \label{eq:sfr_simba}
\end{equation}
where $\epsilon_{*} = 0.02$ is the star formation efficiency, and $t_{\text{dyn}} = 1/\sqrt{G\rho}$ is the local dynamical time. This approach makes star formation dependent on the molecular hydrogen content, which is calculated based on the metallicity and column density of each gas particle using the \citet{KrumholzGnedin2011} sub-grid model. Stellar feedback, i.e. energy and momentum returned to the interstellar medium from supernovae and young stars, is modelled by kinetic outflows, or winds, injected into the surrounding gas. The mass loading factor, $\eta(M_{})$, which scales the wind mass outflow rate relative to the galaxy’s stellar mass, follows a broken power-law based on the Feedback in Realistic Environments (FIRE) simulation data \citep{Muratov2015,Angles-Alcazar2017}, so feedback strength varies by galaxy mass. \simba~also includes a dual-mode black hole growth model, which distinguishes between torque-limited accretion \citep{HopkinsQ2011,Angles-Alcazar2013,Angles-Alcazar2015} for cold gas and Bondi accretion for hot gas \citep{1944MNRAS.104..273B}.

A unique aspect of \simba~is its implementation of AGN feedback, which is the process by which black holes influence their surrounding environment. AGN feedback is split into X-ray feedback plus two modes, dependent on the black hole’s Eddington ratio, $f_{\text{Edd}}$:\\
- Radiative mode (high $f_{\text{Edd}}$), where the black hole ejects gas at moderate speeds (around 1000 km/s), driving molecular and ionised outflows.\\
- Jet mode (low $f_{\text{Edd}}$), where black hole jets drive high-speed outflows in bipolar directions. In this mode, \simba~models the feedback as kinetic energy directed along the black hole’s angular momentum vector, creating jets that can prevent gas from cooling and accreting, effectively quenching star formation in massive galaxies.

Finally, \simba~models dust production and evolution on the fly, accounting for production by stars, growth through metal condensation in the interstellar medium, and destruction from processes like sputtering and supernova shocks. This results in dynamic dust-to-gas and dust-to-metal ratios across environments, which helps accurately predict the mass-metallicity relation without arbitrary adjustments.

Within each snapshot of the simulation, halos and galaxies (central and satellites) are identified with the \caesar~package, a python-based \yt~extension \citep{2011ApJS..192....9T}. Haloes are identified using a standard friends-of-friends (FoF) algorithm with a linking length set to 20\% of the average distance between all particles. To identify galaxies within these haloes, the same FoF method is applied, but only to gas and star particles, and using a much smaller linking length of 0.56\%. This tighter threshold ensures that only particles in dense, gravitationally bound regions, corresponding to galaxies, are grouped together. Once haloes and galaxies are identified, their global properties (e.g. stellar mass, gas content, SFR) are computed by summing the contributions from their associated particles.

In this work, we use the main \simba~runs, which evolve a comoving volume of $100 \, h^{-1}$ cMpc$^{3}$ from $z=256$ to $z=0$. Our analysis focuses on full-physics snapshots at $z=0$, 1, 2, and 3 (or the closest available redshifts), using galaxy properties extracted with \caesar. To investigate recent merger histories, we trace back the most massive progenitors of each galaxy to the preceding snapshot corresponding to a time step of $\Delta t \sim 240$ Myr which corresponds to the time step between the snapshots at $z=0$.

Finally, we also study three ``no feedback'' variant runs at these four redshifts, which have a comoving volume of $50 \, h^{-1}$ cMpc$^{3}$ and the same resolution as the main runs. These correspond to runs i) without X-ray feedback (``no X-ray''), ii) without X-ray, jet, and radiative feedback, i.e. no AGN feedback (``no AGN''), and iii) without AGN and stellar feedback (``no AGN+stellar'').

\subsection{Selection of star-forming galaxies}\label{ssec:selection}

\begin{figure}[h]
        \centering
        \includegraphics[width=\columnwidth]{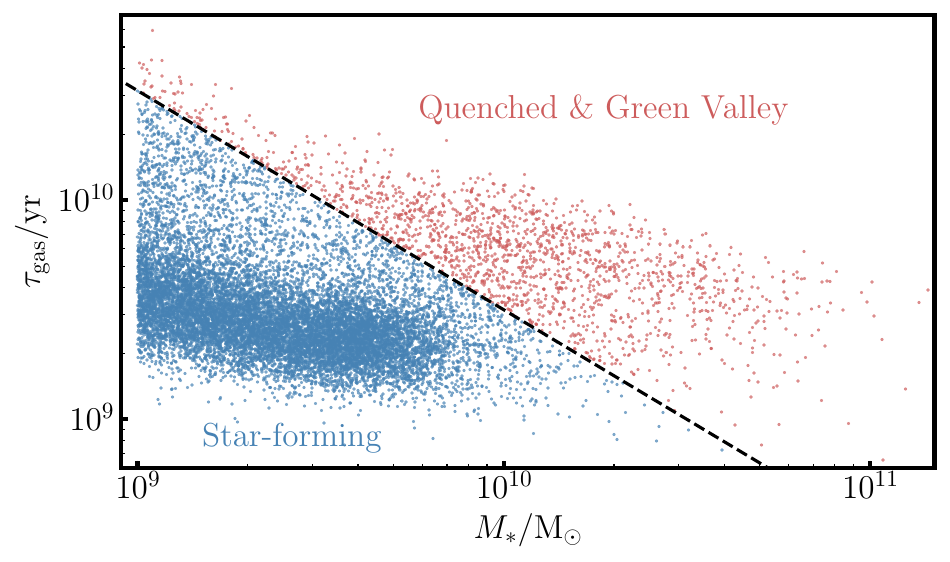}
        \caption{Separation of star-forming galaxies (blue dots) and green valley and quenched galaxies (red dots) using their locus in the $\tau_{\text{gas}} - M_{\star}$ diagram at $z=0$. The selection of $\text{sSFR} > 10^{-10 + 0.3z}$ yr$^{-1}$ has already been applied. Similar cuts are applied at $z=1, 2 \text{ and } 3$ where the fraction of quenched or transitioning galaxies is lower. This selection removes galaxies forming stars slowly compared to their gas mass reservoir.}
        \label{fig:tcut}
\end{figure}

\begin{table}
    \caption{Number of galaxies by type in the main \simba~runs at each selected redshift.}
    \label{tab:simgals}
    \centering
    \begin{tabular}{c c c c c c}
        \hline\hline
        $z$ & Galaxies & Galaxies ($M_{\star} > 10^{9}M_{\odot}$) & SF & GV \& Q \\ 
        \hline
        0 & 55609 & 35060 & 13488 & 19864 \\  
        1 & 39298 & 20792 & 13965 & 4547 \\  
        2 & 30743 & 12240 & 10950 & 626 \\
        3 & 24546 & 8118 & 7868  & 124 \\
        \hline
    \end{tabular}
    \tablefoot{Galaxy types are star-forming (SF), including all main sequence and starbursts galaxies, and green-valley and quenched (GV \& Q). The total number of galaxies is given before and after the low-mass cut, and the numbers for each specific types are given as used here, after the low-mass cut.}
\end{table}

Since this study focuses specifically on the properties of star-forming galaxies, it is essential to adopt a robust selection criterion to ensure a clear and consistent sample across cosmic time. Firstly, we request that each galaxy is composed of at least 50 baryonic particles, which corresponds to a low-stellar-mass limit of $10^{9} \,\text{M}_{\odot}$ in \simba. Then, star-forming galaxies are identified with two complementary selections on the sSFR and on the gas depletion timescale $\tau_{\text{gas}}$ computed as the mass of gas over the SFR.

The first selection corresponds to a lower limit $\text{sSFR} > 10^{-10 + 0.3z}$ yr$^{-1}$, accounting for the changes in amplitude of the main-sequence sSFR with redshift and yielding $\text{sSFR} > 10^{-10}$ yr$^{-1}$ at $z=0$, corresponding to a cut in the colour-colour plane in observations \citep[see e.g.][]{kraljic2018} and a modification of the separation prescribed by \cite{Dave2019} which uses $\text{sSFR} > 10^{-10.8 + 0.3z}$ yr$^{-1}$. This latter difference is intentional, as \cite{Dave2019} aims to distinguish the quenched galaxies from the rest of the population, while we also want to exclude the higher-sSFR part of the green valley and to keep only galaxies that are strictly star-forming.

We complement the sSFR criterion with a second cut based on $\tau_{\text{gas}}$. This quantity is defined as the ratio of a galaxy's gas mass to its current star formation rate, $\tau_{\text{gas}} = M_{\text{gas}}/\text{SFR}$. Physically, it estimates how long the galaxy could sustain its present star formation activity before depleting its gas reservoir, assuming that there is no inflowing or outflowing gas and no merger event. Galaxies with unusually long depletion timescales are typically inefficient at forming stars despite having gas, and may be transitioning toward quiescence. These systems often fall in the green valley and can have sSFR values close to the star-forming threshold, making them ambiguous cases. To remove such outliers, we apply an additional criterion $\tau_{\text{gas}} < 10^{19.5 - \log(M_{\star}/\text{M}_{\odot})}$ yr. This corresponds to a separation between two populations in the $\tau_{\text{gas}} - M_{\star}$ plane as shown in Fig.~\ref{fig:tcut} at $z=0$, but the exact value and slope of this cut are not crucial as long as it stays above the main star-forming population (dense part of the scatter), as it removes only a very marginal number of galaxies with relatively high depletion timescale for their masses. The number of each type of galaxies at the selected redshifts is given in Table~\ref{tab:simgals}.

In Sect.~\ref{sec:results}, we discuss the dependency of our galaxy population on the cuts on sSFR and $\tau_{\text{gas}}$ at $z=0$ as it corresponds to the epoch at which galaxies had time to quench or to enter a transition phase through the green valley.

\section{Methods}\label{sec:methods}

\begin{figure*}
        \centering
        \includegraphics[width=\textwidth]{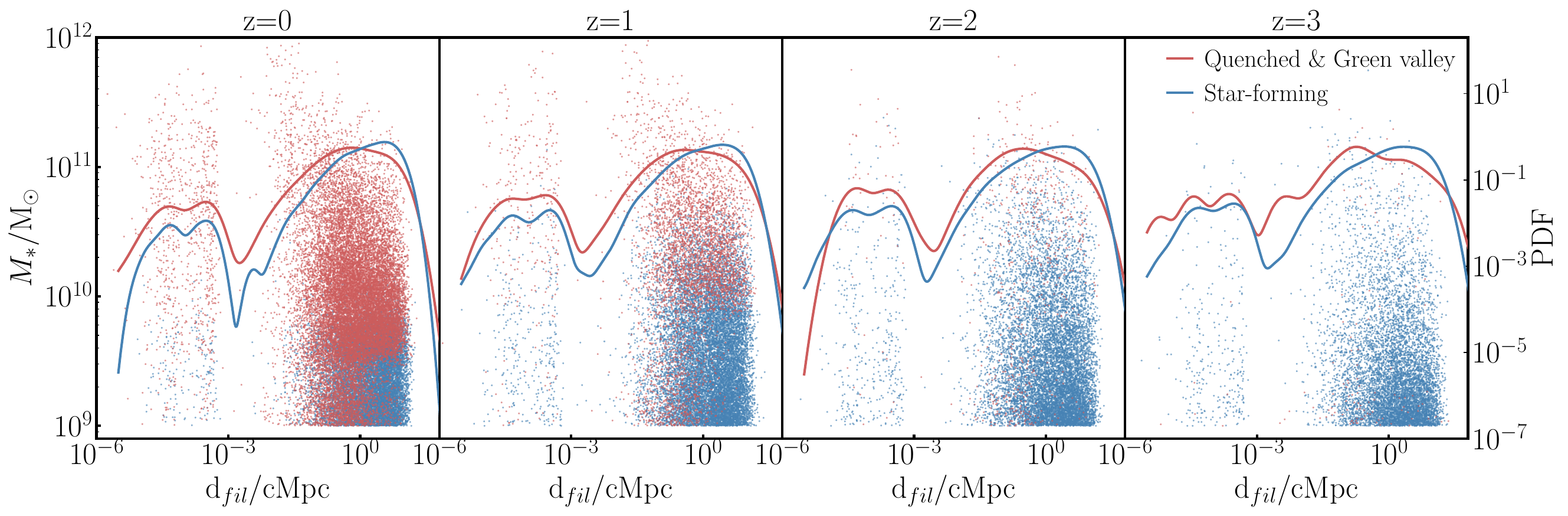}
        \caption{Stellar mass of star-forming galaxies (blue) and quenched and green-valley galaxies (red) in \simba~at $z=0$, 1, and 2 (from left to right) as a function of the distance to the closest filament. Each dot represents a galaxy position in the stellar mass-$\mathrm{d}_{fil}$ parameter space (left side y-axis), and the solid lines are the corresponding smoothed and normalised distributions of distances for each population (right side y-axis). The visual inspection reveals the emergence of quenching at $z\sim1$ and only a marginal quenched galaxy population a $z\geq2$, with massive quenched galaxies lying typically closer to filaments than star-forming galaxies at all redshifts with a significant quenched population at $z=0$.}
        \label{fig:distrib}
\end{figure*}

In this section, we explain how we derive 3D cosmic web skeletons from our simulated data, which are essential for quantifying the spatial relationship between galaxies and the filamentary large-scale structure. By constructing a cosmic web skeleton, we can measure each galaxy's distance to the nearest filament, providing a basis for studying how galaxy properties relate to their position within the cosmic web.

\subsection{Cosmic Web skeletons from \disperse}\label{ssec:disperse}

Within each snapshot, we use publicly available software \disperse\footnote{\href{https://www2.iap.fr/users/sousbie/web/html/indexd41d.html}{https://www2.iap.fr/users/sousbie/web/html/indexd41d.html}} \citep{Sousbie2011, SousbiePK2011}, to extract the cosmic web skeleton from the positions of the entire population of galaxies. \disperse~uses discrete Morse theory and allows us to reconstruct the skeleton of the cosmic web, composed of critical points, from the density field gradient computed via Delaunay tessellation \citep{Schaap2000}. Each filament is a set of segments connecting the extrema.

We apply \disperse~to extract the cosmic web skeleton from the distribution of all galaxies with stellar mass $M_{\star} > 10^{9}\rm M_{\odot}$ in \simba, using a persistence threshold of 3$\sigma$, a commonly adopted value in similar analyses \citep[e.g.][]{malavasi2017, kraljic2018, Sarron2019}. We choose this mass cut for consistency, but we have verified that including all galaxies does not change our analysis and results. In addition, we apply a single smoothing step to the filament skeleton, which averages the positions of adjacent segment endpoints to eliminate small-scale noise and non-physical sharp bends. Finally, we define the distance from the filament, denoted $\mathrm{d}_{fil}$, as the shortest 3D distance from a galaxy to the closest filament, computed via orthogonal projection using a k-d tree algorithm \citep{maneewongvatana1999}.

\subsection{Selection of galaxy properties}\label{ssec:quantities}

In the rest of this paper, we focus on the star-forming galaxies at $z=0, 1, 2 \text{ and } 3$ selected as presented in Sect.~\ref{sec:data}.

Firstly, we want to examine the dependence of galaxy stellar masses on $\mathrm{d}_{fil}$, the distance to the closest filament. Observational studies suggest that the median mass of galaxies tends to increase with their proximity to the closest filament or node. This is consistent with the idea that environmental effects on halo masses drive the most massive halos toward filaments, resulting in a higher concentration of massive galaxies in these regions \citep[e.g.][]{malavasi2017, kraljic2018, laigle2018}. This trend is generally attributed to the modulation of halo and galaxy growth by the large-scale environment, as first discussed in the context of density fluctuations by \cite{Kaiser1984}, as gravitational interactions and accretion processes facilitate the growth of galaxies in dense environments. However, recent simulation results have revealed important subtleties. For example, \cite{10.1093/mnras/stae667} showed that while the overall trend of increasing stellar mass toward filaments is reproduced for central galaxies, satellite galaxies can exhibit the opposite behaviour in \simba~at $z=0$ and $z=1$. Satellites closer to filaments may have lower stellar masses, possibly due to environmental quenching and stripping processes. Notably, most observational studies to date do not distinguish between centrals and satellites, which can mask these differential trends. Understanding this mass distribution is essential, as stellar mass strongly influences various galaxy properties, including star formation rate, gas content, and metal enrichment. Given this, it is necessary to carefully account for mass effects when analysing the evolution of star-forming galaxies within the cosmic web.

The mass distribution of star-forming, and quenched and green-valley galaxies as a function of $\mathrm{d}_{fil}$ in Fig.~\ref{fig:distrib} reveals a pronounced gap in the intermediate range $10^{-3} \lesssim \mathrm{d}_{fil}\lesssim 10^{-2}$ cMpc as the vast majority of galaxies populate a broad sequence from the filaments out to about ten comoving Mpc (cMpc), and a small number of objects lie closer than d$_{fil} = 10^{-3}$ cMpc, typically in massive haloes near nodes Increasing the skeleton smoothing from one to two operations leaves this pattern unchanged. As detailed in Appendix~\ref{appA}, raising the \disperse~persistence threshold from our fiducial $3\sigma$ to $5\sigma$ reduces the counts at the smallest separations which makes the contrast with the intermediate‐distance deficit less striking, but it does not change the results presented in our analysis.

In the following, we include all distance ranges in the analysis but treat d$_{fil}<10^{-2}$ cMpc as a separate innermost bin, and we focus primarily on galaxies at d$_{fil} > 10^{-2}$ cMpc, which is by far the most numerous sample, to characterise how galaxy properties evolve with proximity to the cosmic web. To do so, galaxies at the smallest distances are grouped in a separate single distance bin. In addition, in order to account for the stellar mass dependence of many galaxy properties considered in this work, we proceed as follows. For each dimensionless quantity $Q$, we study the evolution of its mean deviation from its stellar-mass-dependent best-fit double power-law $Q_{bf}$, computing $\Delta \log(Q)$ in bins of distance to the filament as
\begin{equation}
    \Delta \log(Q)_{\text{bin}} = \frac{\sum_{i=1}^{N_{\text{gal}}}\log\left(\frac{Q_{i}}{Q_{i, bf}} \right)}{N_{\text{gal}}},
    \label{eq:Dlog}
\end{equation}
where we sum over the number of galaxies in the bin $N_{\text{gal}}$. In practice, for each galaxy, we compute the desired quantity and divide it by the best-fit value at this galaxy mass. We adopt a double power-law 
\begin{equation}
    Q_{bf}(M_{\star}) = \frac{A}{(\frac{M_{\star}}{M_{\text{break}}})^{\alpha} + (\frac{M_{\text{break}}}{M_{\star}})^{\beta}}
    \label{eq:Qbf}
\end{equation}
with the amplitude $A$, the mass at slope break $M_{\text{break}}$, and the slopes $\alpha$ and $\beta$ as fitted parameters. This double power-law provides a slightly better fit than a single power-law and avoids the overfitting produced by a running median. In practice, we first determine for each physical quantity its best-fit double power-law relation with $M_{\star}$, fitted over the entire star-forming population at each redshift. Each individual measurement $Q_{i}$ is then compared to its best-fit value $Q_{\mathrm{bf}}(M_{\star})$. This procedure allows us to evaluate trends of galaxy properties with filament proximity beyond their underlying stellar mass dependence. For each quantity, we also compute the median deviation from the best-fit relation in each distance bin.

In order to examine the behaviour of star-forming galaxies within the cosmic web, we primarily study their SFR, sSFR and molecular hydrogen depletion timescale $\tau_{\rm H_{2}} = M_{\rm H_{\text{2}}}/\text{SFR}$. Here, $M_{\rm H_{\text{2}}}$ refers to the molecular hydrogen mass tracked in the simulation, and the SFR is measured over the local dynamical time $t_{\text{dyn}}$. Consequently, $\tau_{\rm H_{\text{2}}}$ represents the timescale over which the current molecular gas reservoir would be depleted if the SFR remained constant. These quantities are expected to be connected since the SFR given in \simba~is computed from the molecular gas density, and $\tau_{\rm H_{\text{2}}}$ characterises the timescale of conversion of the molecular gas into stars, driving the SFR. In addition, we also study the gas depletion timescale $\tau_{\text{gas}} = M_{\text{gas}}/\text{SFR}$ where $M_{\text{gas}}$ is the total gas mass, the total gas mass fraction calculated as $M_{\text{gas}}/(M_{\star} + M_{\text{gas}})$, and the molecular-to-atomic hydrogen mass fraction $M_{\rm H_{\text{2}}}/(M_{\ion{H}{i}} + M_{\rm H_{\text{2}}})$. These fractions quantify the gas reservoirs relative to the baryonic masses of the galaxies and the phase of this gas, and thus provide insight into how efficiently galaxies replenish and consume their gas over time. They also help us understand the role of environmental processes in regulating star formation by tracking the availability of different gas phases within the cosmic web.

\section{Results}\label{sec:results}

In this section, we examine the residuals of different galaxy properties (Eq.~\ref{eq:Dlog}) as a function of the proximity of galaxies to filaments $\mathrm{d}_{fil}$. We focus on the entire star-forming galaxy population at $z=0, 1, 2 \text{ and } 3$. All statistical errors are computed by bootstrapping over a hundred random sub-samples of the parent sample with replacement.

\subsection{Galaxy properties as a function of their distance to filaments}\label{ssec:results}

\begin{figure*}
    \centering
    \includegraphics[width=\textwidth]{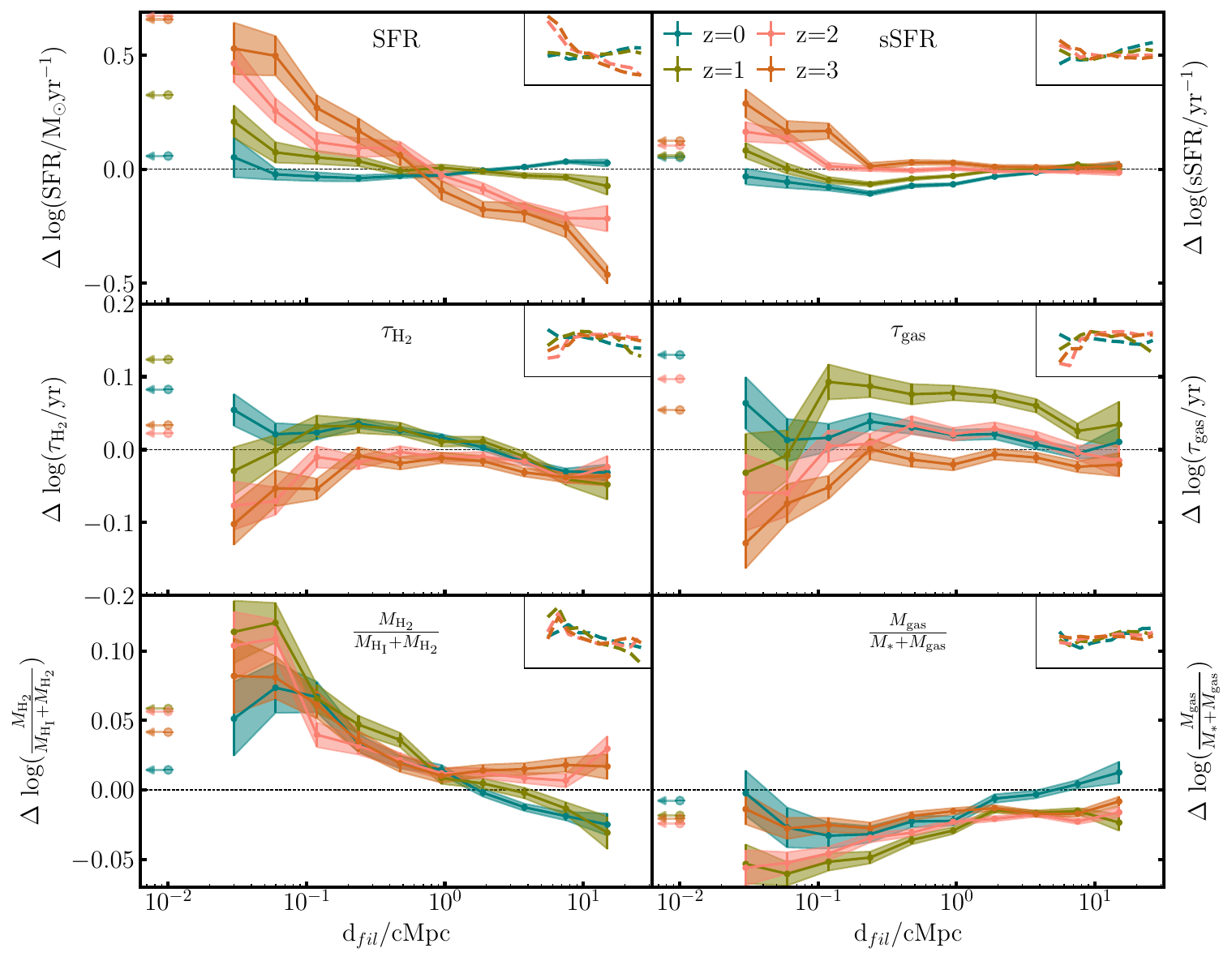}
    \caption{Residuals deviation as a function of the distance to the filaments after correcting for stellar-mass effects. We show the SFR (top left), the sSFR (top right), the molecular hydrogen depletion timescale (middle left), the total gas depletion timescale (middle right), the molecular hydrogen fraction (bottom left), and the gas fraction (bottom right) for star-forming galaxies in \simba~in bins of distance at $z=0$ (blue), 1 (green), 2 (red), and 3 (brown). The arrows pointing to the left represent galaxies located at smaller distances than the plotted range, grouped into a single bin. For each panel, the upper right box shows the deviation of the in-bin median from the best-fit relation. Star formation and gas properties vary systematically with filament proximity with enhancement near filaments at high z, but a V-shaped trend in sSFR and $\tau_{\rm H_{2}}$ at $z=0$.}
    \label{fig:DeltaQ_arranged}
\end{figure*}

\begin{figure}[h]
    \centering
    \includegraphics[width=\columnwidth]{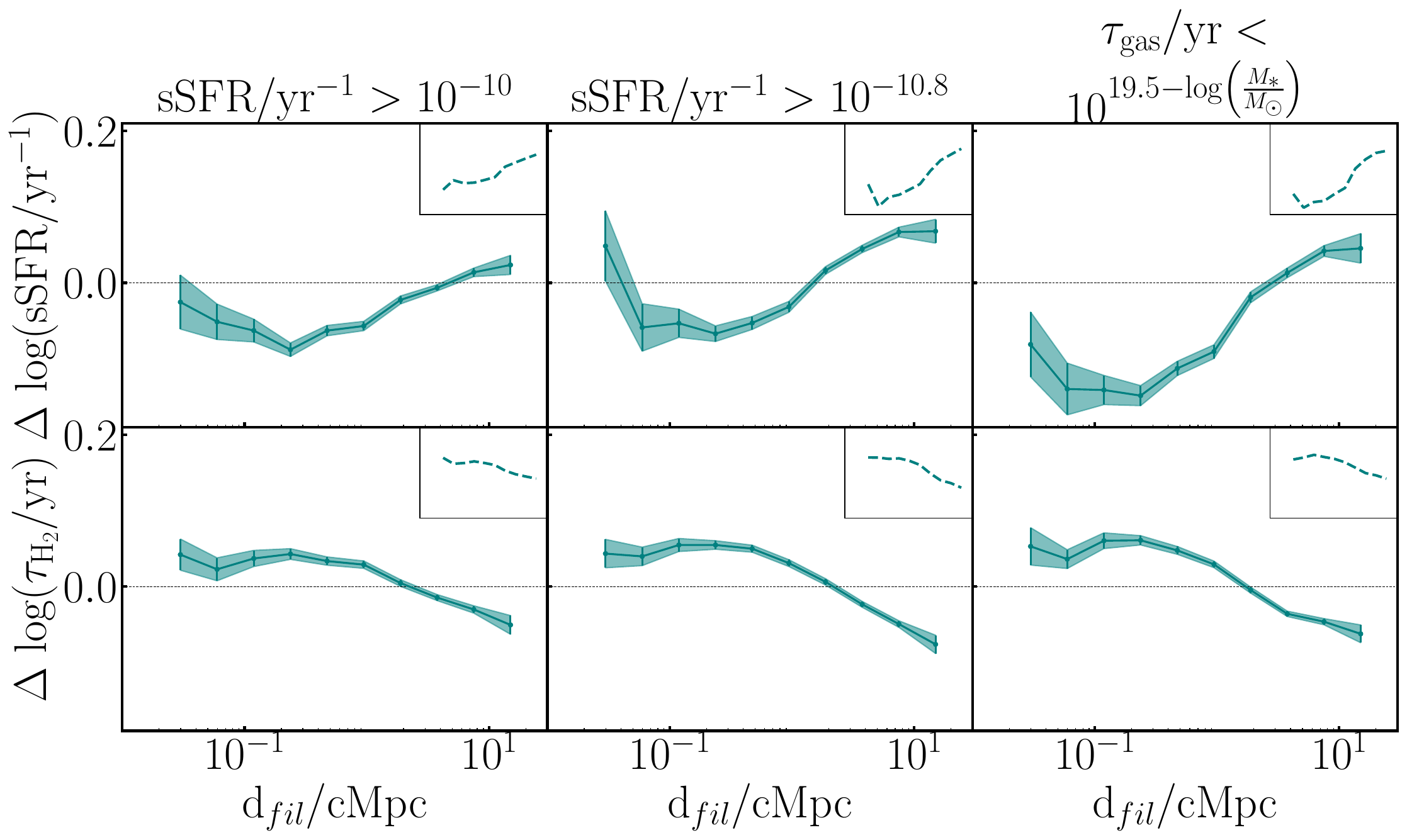}
    \caption{Deviation of the mean specific star-formation rate (top) and molecular hydrogen depletion timescale (bottom) from the respective best-fit relations in logarithmic scale as a function of $d_{\rm fil}$ in \simba~at $z=0$, using different selection criteria for star-forming galaxies. For each panel, the upper right box shows the deviation of the in-bin median from the best-fit relation. The qualitative sSFR and $\tau_{\rm H_{2}}$ trends with filament distance at z = 0 are robust against variations in the star-forming selection criteria.}
    \label{fig:selection}
\end{figure}

Fig.~\ref{fig:DeltaQ_arranged} shows the mean deviations from the stellar-mass-dependent best-fit relations for SFR, sSFR, $\tau_{\rm H_{2}}$, $\tau_{\rm gas}$, the molecular gas mass fraction, and the total gas mass fraction, in bins of distance to the nearest filament, at the four redshifts of interest, with all the galaxies at d$_{fil} < 10^{-2}$ cMpc in one single separate bin represented by a left-pointing arrow. At $z=2$–3, SFR increases steadily as galaxies get closer to filaments. The sSFR exhibits a similar behaviour, although its profile flattens at d$_{fil} \gtrsim 0.25$ cMpc. At $z=1$, the sSFR shows a shallow minimum near d$_{fil} \sim 0.25$ cMpc. At $z=0$, both SFR and sSFR follow a clear V-shaped trend, with a minimum at intermediate distances and a rise in the innermost bin (d$_{fil} < 0.25$ cMpc). When quenched galaxies, which dominate at $z < 1$, are included in the sample, this upturn disappears, and the overall trend becomes a decline of sSFR toward filaments in agreement with previous findings of \cite{10.1093/mnras/stae667}. The molecular gas depletion timescale $\tau_{\rm H_{2}}$ follows an inverse trend to the SFR at all redshifts, decreasing toward filaments, except at $z=0$ where $\tau_{\rm H_{2}}$ increases in the innermost distance bin. The molecular gas fraction increases steadily with proximity to filaments at $z \geq 1$, peaking in the closest distance bin. At $z=0$, however, the molecular gas fraction drops by approximately 0.2 dex in the innermost bin. The total gas fraction decreases with decreasing distance at $z \geq 1$, but shows a V-shaped trend at $z=0$.

We verified that these trends are robust against variations in the definition of star-forming galaxies. A representative example of these tests at $z=0$ is shown in Fig.~\ref{fig:selection}, where we tested three alternative criteria with a pure sSFR cut of $\text{sSFR} > 10^{-10}$ yr$^{-1}$, a less restrictive sSFR threshold of $\text{sSFR} > 10^{-10.8}$ yr$^{-1}$, and a cut only on the gas depletion timescale defined as $\tau_{\rm gas} < 10^{19.5 - \log(M_{\star}/\rm M_{\odot})}$ yr. All of these alternative selections yield qualitatively similar trends. Our main criterion, described in Sect.~\ref{ssec:selection}, combines an sSFR threshold of $10^{-10 + 0.3z}$ yr$^{-1}$ with the additional depletion timescale cut to ensure a conservative and consistent selection of strictly star-forming galaxies across redshifts. At higher redshifts ($z \geq 1$), where the quenched fraction is much lower, the choice of selection has a negligible impact on the results.

To further understand the processes underlying the observed redshift-dependent trends in star formation and gas properties as a function of filament proximity, we next examine potential physical drivers. Specifically, we analyse the recent merging history of galaxies over the past 240 Myr, the dichotomy between central and satellite galaxies given their distinct environmental interactions, the impact of feedback processes as implemented in the \simba~simulation, and the gas accretion rates within the same recent time interval. This approach aims to disentangle the relative contributions of these factors to the modulation of star formation in star-forming galaxies, with a focus on the trends in sSFR and SFR at $z=0$. We note that as behaviours are similar at $z=2$ and $3$, we only show the results up to $z=2$ to simplify the figures.

\subsection{Effect of mergers}\label{ssec:mergers}

\begin{figure}
        \centering
        \includegraphics[width=\columnwidth]{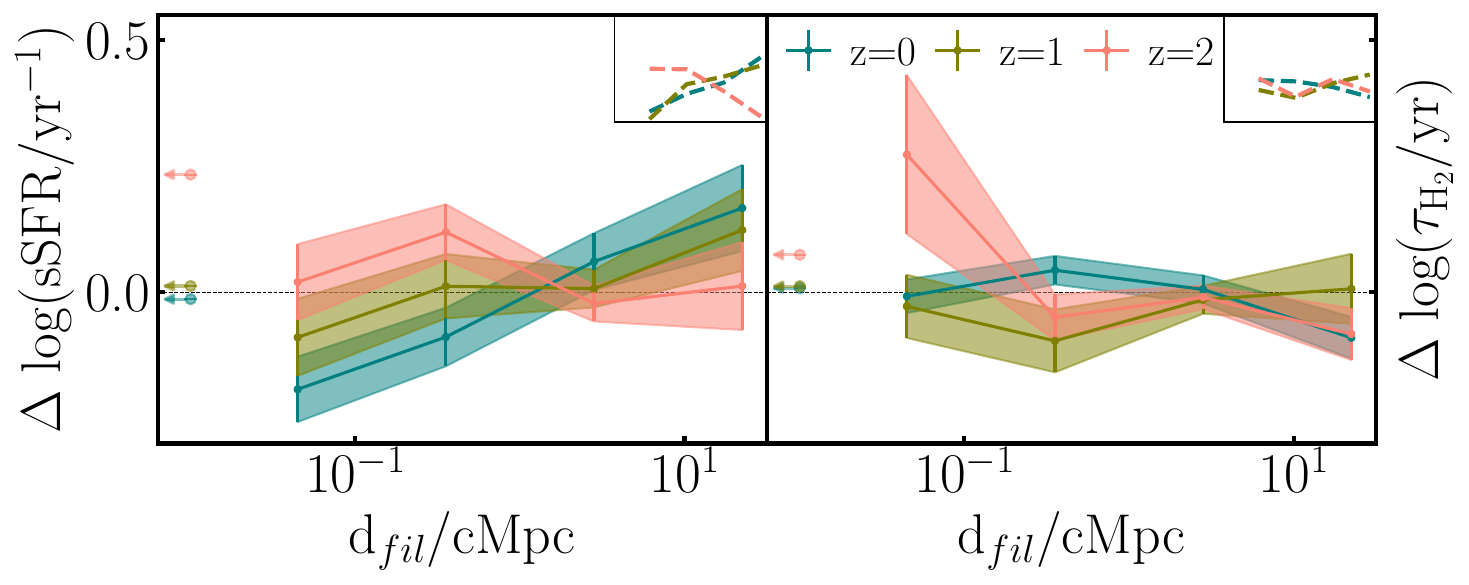}
        \caption{Same as Fig.~\ref{fig:DeltaQ_arranged} only for galaxies which underwent a recent major merging event, for sSFR and $\tau_{H_{2}}$ only. Star-forming galaxies with recent major mergers do not show the V-shaped sSFR trend, suggesting mergers are not the primary driver of environmental modulation.}
        \label{fig:DeltaQ_mergers}
\end{figure}

\begin{figure}
        \centering
        \includegraphics[width=\columnwidth]{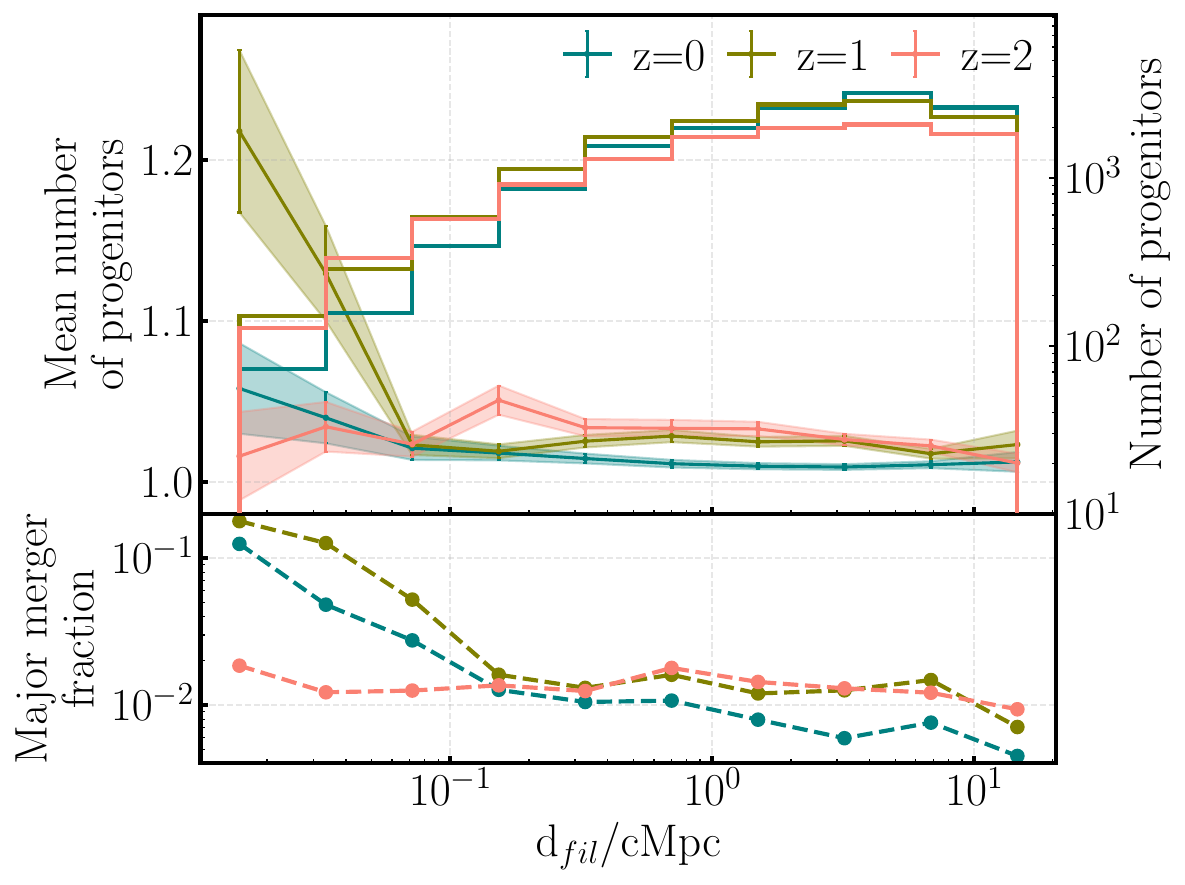}
        \caption{Top panel: mean number of progenitors per galaxy and distribution of the number of progenitors as a function of the distance to the closest filament at $z=0$, 1 (green), and 2 (red). Bottom panel: fraction of major mergers, defined as galaxies with secondary progenitors with a stellar-mass ratio of at least 1:4, for the three redshifts as a function of the distance to the closest filament. The frequency of recent mergers increases near filaments at low redshift, but major mergers remain rare and are unlikely to dominate environmental trends.}
        \label{fig:merger_combined}
\end{figure}

To investigate the role of mergers in modulating star formation activity as a function of filament proximity, we analyse the merger history of star-forming galaxies between consecutive snapshots, thus over the past $\sim$240 Myr. This fixed interval corresponds to the time separation between two consecutive snapshots at $z=0$, and is used consistently at all redshifts to ensure a uniform merger selection timescale. This choice is consistent with previous studies showing that the peak SFR enhancement from major mergers occurs within 100-250 Myr following the event \citep[e.g.][]{DiMatteo2007}. Since filaments are dense, dynamically active regions that channel matter and galaxies into nodes, one might naively expect a higher merger rate and consequently more merger-driven starburst activity near filaments. Fig.~\ref{fig:DeltaQ_mergers} presents the evolution of sSFR and $\tau_{H_{2}}$ as a function of $\mathrm{d}_{fil}$ for galaxies that have undergone major merger events, defined as merging events where the secondary progenitor galaxy has at least 25\% of the stellar mass of the resulting galaxy. We verified that SFR shows a similar behaviour as sSFR in these galaxies and comment on this in the following. The results reveal a striking contrast to the trends observed in the full star-forming population. For galaxies with major mergers, the V-shaped sSFR-$\mathrm{d}_{fil}$ relation at $z=0$ disappears. Instead, both sSFR and SFR decrease monotonically with decreasing $\mathrm{d}_{fil}$, particularly toward filament cores. At $z=1$, we observe a flattening of the sSFR profile toward filaments, while at $z=2$, the enhancement of sSFR at a small $\mathrm{d}_{fil}$ remains, although less pronounced compared to the entire population. The H$_2$ depletion timescale is relatively flat at all $z$, with a slight increase in the innermost bin at $z=2$. This remains true even when changing the definition of a major merger, for instance by requiring the secondary progenitor galaxy stellar-mass to be at least 33\% or 20\% the stellar-mass of the resulting galaxy, as both yield qualitatively identical results. The central values may shift slightly, yet always within the corresponding uncertainties. The main impact is on the statistical error bars, with slightly larger error bars for stricter (20\%) definitions due to smaller sample sizes, and smaller error bars for less strict definitions. However, these results should be interpreted with caution, particularly at $z=0$, where the limited number of major mergers among star-forming galaxies restricts our analysis to only four distance bins within the distance range of interest. Based on the literature, we also note that mergers with smaller mass ratios, typically with secondary progenitors with stellar masses below 20\% that of the resulting galaxy, produce little to no measurable impact on SFR over long timescales \citep[e.g.][]{Cox2008}. In addition, we verify that changing the time window from $\sim240\ \rm Myr$ to $\sim500\ \rm Myr$ at $z=0$ does not help recover V-shaped $\Delta \log(\rm sSFR/yr^{-1})$ as a function of $\mathrm{d}_{fil}$, although it roughly doubles the number of detected major mergers. Overall, these tests confirm that our results are robust against reasonable variations in both the major-merger definition and the temporal window considered.

Figure~\ref{fig:merger_combined} provides additional insights for interpreting these trends. The upper panel shows the mean number of progenitors per galaxy, computed over the past $\sim$240 Myr, as a function of $\mathrm{d}_{fil}$. This number is one in the absence of mergers and a higher value indicates recent merging events. At $z=0$ and $z=1$, the mean number of progenitors increases as galaxies approach filaments, reaching a plateau at the largest distances. In contrast, at $z=2$, the mean number of progenitors remains relatively flat in all distance bins, indicating a weaker environmental dependence of merger activity at higher redshift. The fraction of galaxies that have experienced major mergers, displayed in the lower panel, follows a similar pattern as it increases toward filaments at $z \leq 1$, while remaining relatively uniform at $z=2$. These findings suggest that while mergers become more frequent in filamentary environments at lower redshifts, there is no evidence that major mergers are responsible for the V-shaped sSFR-$\mathrm{d}_{fil}$ relation observed at $z=0$ in the full star-forming sample, and the overall contribution of major mergers to observed trends is likely subdominant. Moreover, the lack of any enhancement in merger activity near filaments at $z \geq 2$ strongly indicates that mergers cannot account for the rise of SFR in these environments at high redshift, where this increase is even more pronounced than the V-shaped pattern seen at $z=0$.

\subsection{Central and satellite galaxies}\label{ssec:c_vs_s}

\begin{figure}
        \centering
        \includegraphics[width=\columnwidth]{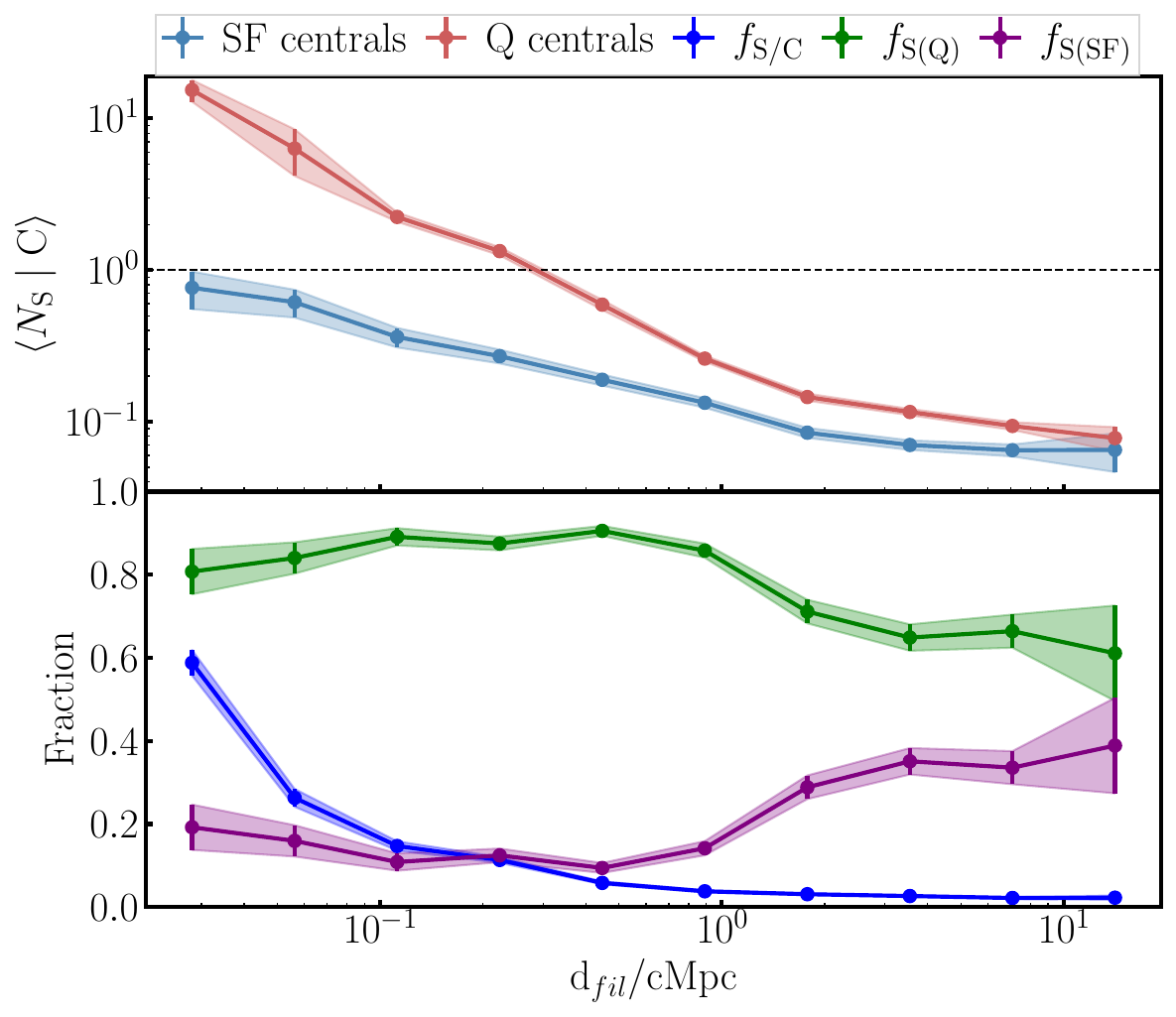}
        \caption{Top panel: Mean number of satellites per central galaxy as a function of distance to the nearest filament, shown separately for star-forming (blue) and quenched (red) centrals. Both populations show a steep rise in $N_{\mathrm{sat}}$ towards filament cores. Bottom panel: Fraction of star-forming satellites among star-forming galaxies (blue), fraction of satellites of quenched centrals within star-forming satellites (green), and satellites of star-forming centrals (purple). The satellite fraction increase towards the filaments core and are dominated everywhere by satellites of quenched centrals.}
        \label{fig:stats}
\end{figure}

\begin{figure}
        \centering
        \includegraphics[width=\columnwidth]{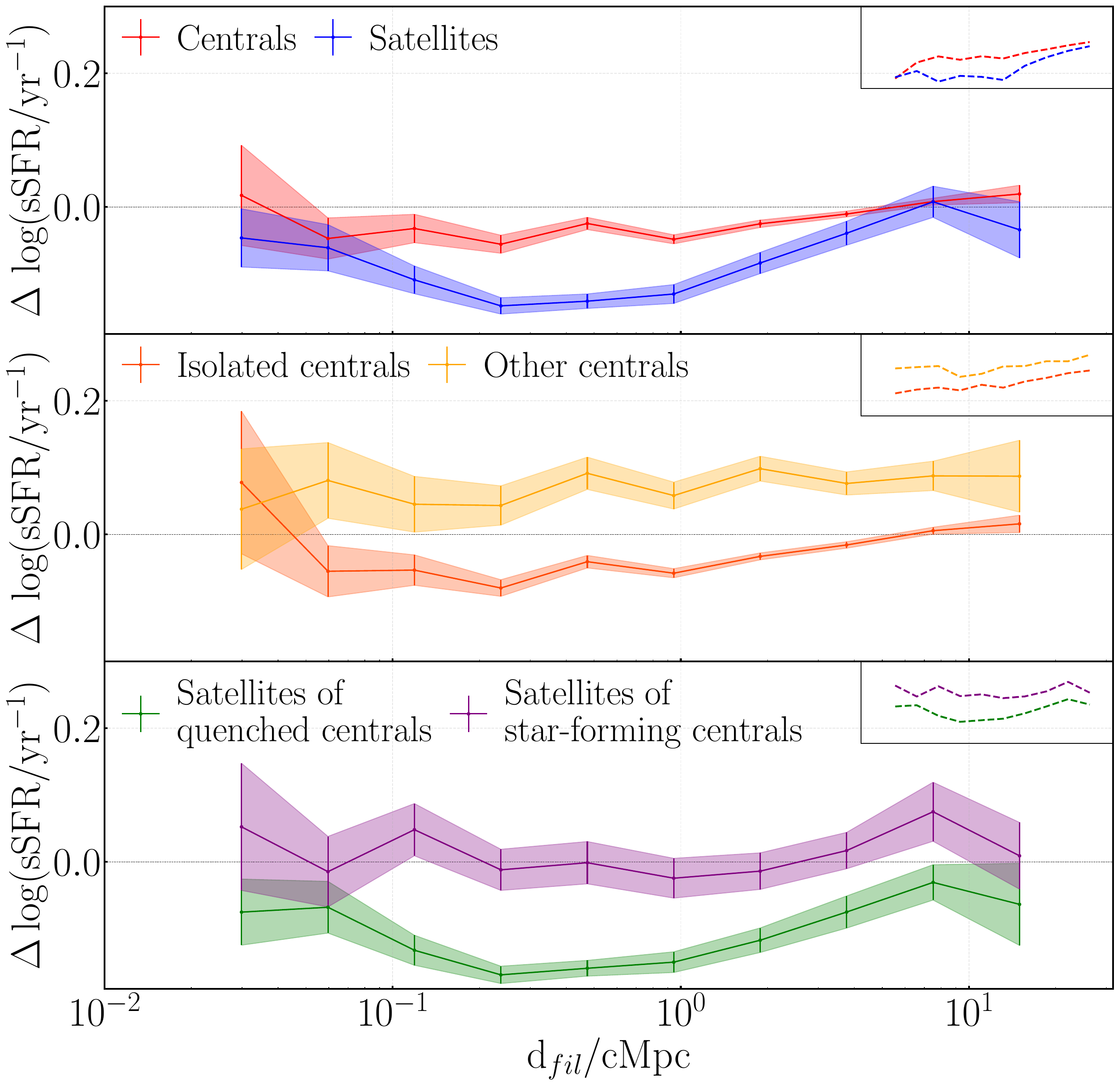}
        \caption{Deviation of the mean specific star-formation rate for central and satellite galaxies (red, blue, top panel), field and cluster central galaxies (orange-red, orange, middle panel), and satellite of quenched central and star-forming central galaxies (green, purple, bottom panel) from the best-fit relation in logarithmic scale as a function of the distance to the closest filament for star-forming galaxies in \simba~in bins of distance at $z=0$. For each panel, the upper right box shows the deviations of the in-bin medians from the best-fit relation. The V-shaped sSFR–filament distance trend at $z=0$ is driven primarily by satellites, particularly those orbiting quenched centrals.}
        \label{fig:distinctionz0}
\end{figure}

We now verify environmental dependencies of central and satellite galaxies, focusing first on the large-scale environment as probed by central galaxies and then on the group- and cluster-scale environment manifesting in satellite galaxies. The analysis in this section is focused on $z=0$ as we aim to understand the causes of the V-shaped trend seen in the sSFR-$\mathrm{d}_{fil}$ relation. Fig.~\ref{fig:stats} summarises key quantitative features of the central–satellite distribution. The upper panel shows the mean number of satellites per central galaxy, measured separately for star-forming (blue) and quenched (red) centrals, with shaded regions indicating $1\sigma$ uncertainties. Star-forming centrals host on average fewer satellites than quenched centrals at all $\mathrm{d}_{fil}$, but in both cases $N_{\mathrm{sat}}$ rises steeply towards the filament core. For star-forming centrals, the mean number exceeds $\sim0.5$ at d$_{fil}\approx0.1$ cMpc, indicative of environments where centrals are embedded in rich local structures. At large distances (d$_{fil}\gtrsim0.5$ cMpc), most centrals are isolated or have only a few satellites. The lower panel shows the fraction of star forming galaxies that are satellites (blue), the fraction of satellites of quenched centrals (green), and satellites of star-forming centrals (purple). The satellite fraction increases continuously towards smaller $\mathrm{d}_{fil}$, exceeding $0.5$ at the smallest $\mathrm{d}_{fil}$, and is dominated everywhere by satellites of quenched hosts, with up to a fraction of 0.8 for d$_{fil}\lesssim1$ cMpc. In the outskirts (d$_{fil}\gtrsim0.5$ cMpc), satellites of quenched centrals still make up a substantial $\sim60\%$ of the satellite population, indicating a more mixed population at large distances. These trends confirm that filament interiors are satellite-rich environments, particularly around quenched centrals, and that the local environment traced by $\mathrm{d}_{fil}$ strongly correlates with the central–satellite makeup of the galaxy population. At this stage, however, we cannot yet determine whether satellites also exhibit suppressed sSFR relative to centrals. Fig.~\ref{fig:distinctionz0} explores the star formation suppression more explicitly by showing the residuals of sSFR across three central-satellite sub-populations. In the upper panel, we compare the full sample of centrals and satellites as a function of $\mathrm{d}_{fil}$. A clear dichotomy emerges as satellites exhibit a pronounced V-shaped dependence, with a significant suppression of sSFR at intermediate distances (d$_{fil} \sim 0.2$–0.4 cMpc), whereas centrals show an almost flat trend, indicating little modulation of their star-forming activity by proximity to the filament core. The shape seen for satellites mirrors that of the global star-forming population, implying that the overall environmental signal is dominated by satellites, particularly in the intermediate regions of filaments (i.e. between the dense cores and lower density outer regions). The middle panel further isolates centrals by environment, distinguishing those with no satellites (isolated centrals) from those embedded in groups (centrals with satellites). Isolated centrals are statistically dominant, and thus their sSFR residuals qualitatively resemble those seen with the full central sample, whereas group centrals show flatter behaviour. This suggests that any effect that satellites might have on their host centrals is therefore diluted in the combined central sample. The imprint of satellites on the signal instead emerges in other parts of the analysis, particularly when examining satellites as a separate population. The bottom panel focuses on satellites alone, restricted to the star-forming population, and splits them by the type of their host central (star-forming or quenched). Star-forming satellites of quenched central galaxies show a very strong V-shaped trend, unlike those around star-forming centrals, and the two populations are inconsistent within 1$\sigma$ across most bins. The enhanced suppression of sSFR among satellites of quenched hosts highlights a form of conformity, reflecting a shared environmental history or intra-halo processes that affect all constituents within the system.

Altogether, the central-satellite dichotomy clarifies that the environmental dependence of sSFR on the position in the cosmic web at $z=0$ is not uniform across all galaxies, but is instead strongly mediated by the environmental substructure and hierarchical position. Satellites are the main drivers of the environmental signal, whereas centrals retain the ability to form stars largely irrespective of their large-scale cosmic location. We note that removing galaxies in massive clusters, i.e. in halos with $M_h > 10^{13} \rm M_{\odot}$, has a negligible effect on our conclusions (see Appendix~\ref{appC}). Therefore, the trends described here are not driven primarily by cluster proximity, but rather by distance to the cosmic web skeleton and the associated galaxy type. Although this section focuses on $z=0$, we note that qualitatively the central-satellite separation remains relevant at higher redshifts. However, the decreasing satellite fraction with increasing redshift reduces the impact of satellite-driven effects on the overall environmental trends, making centrals more representative of the star-forming population beyond $z \sim 1$.

\subsection{Effect of feedback}\label{ssec:feedback}

\begin{figure}
        \centering
        \includegraphics[width=\columnwidth]{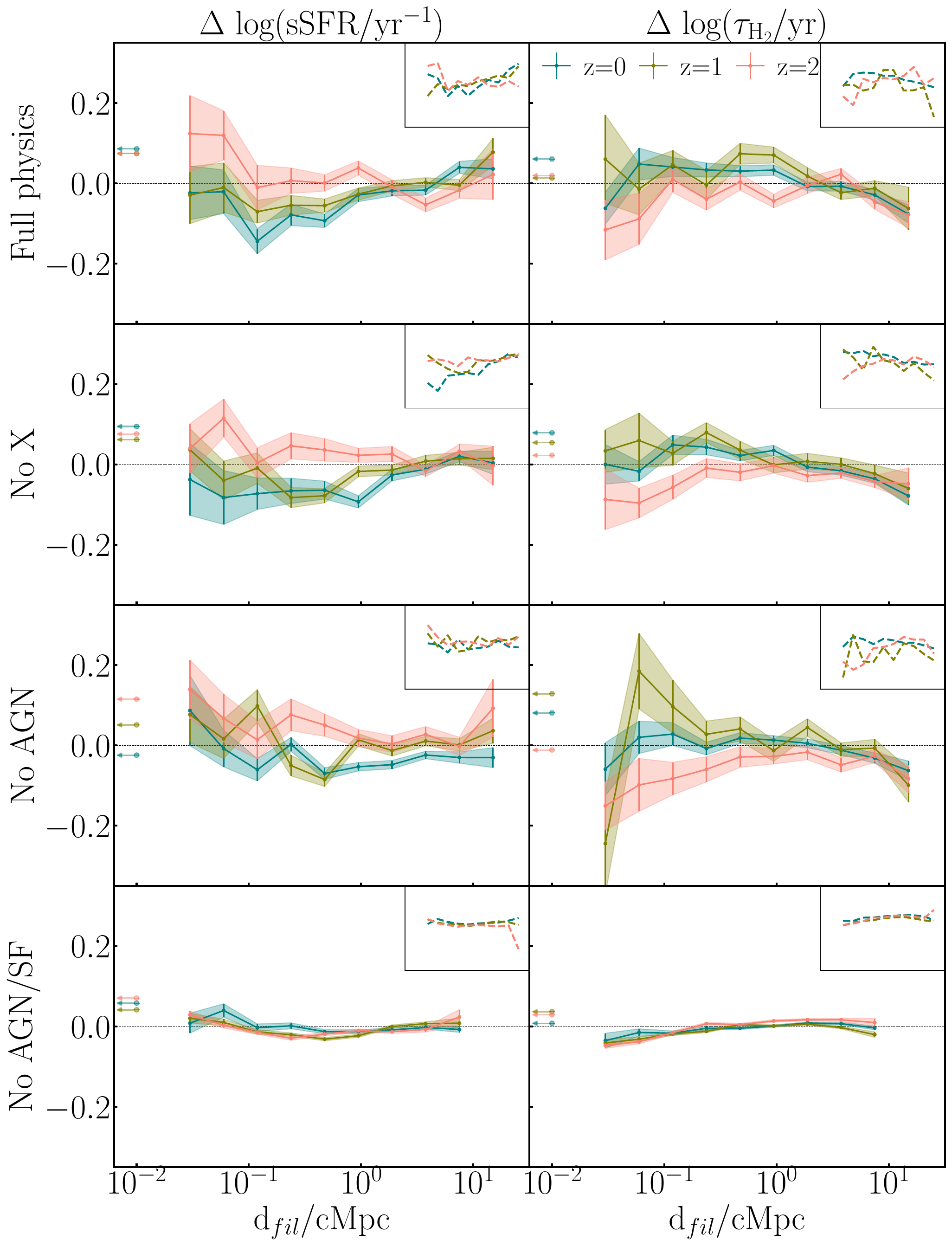}
        \caption{Deviation of the mean specific star-formation rate (top) and molecular hydrogen depletion timescale (bottom) from the respective best-fit relations in logarithmic scale as a function of $\mathrm{d}_{fil}$ in \simba~in bins of distance at $z=0 \text{ (blue) }, 1 \text{ (green), and } 2 \text{ (red)}$. The arrows pointing to the left give the value for a bin containing all the galaxies not included in the distance range we study here. For each quantity, the upper right box shows the deviation of the in-bin median from the best-fit relation. From left to right, the columns give the results for the full physics, no X-ray feedback, no AGN feedback, and no AGN and stellar feedback in the 50cMpc$^{3}$ runs. Removing stellar feedback processes flattens sSFR–distance trends.}
        \label{fig:feedback}
\end{figure}

\simba~implements multi-scale feedback processes that shape galaxy evolution through distinct physical mechanisms. AGN feedback manifests itself through three primary channels: X-ray heating (thermal energy injection into circumgalactic gas), radiative winds (momentum-driven outflows), and relativistic jets (kinetic energy injection along angular momentum axes). Stellar feedback arises from supernovae-driven winds that eject gas from star-forming regions. To disentangle their roles in environmental trends, we analyse four configurations, namely ``full physics'', ``no X-ray'', ``no AGN'' (stellar only), and ``no AGN+stellar'' (no feedback) at $z=0$, 1, and 2 shown in Fig.~\ref{fig:feedback} for the entire population of galaxies. We also repeated this analysis for satellites and centrals alone, but with the lack of statistics in the smaller \simba~boxes, there is no evidence of clear differences between these populations. To guide interpretation, we start from the top panels and progressively remove individual feedback mechanisms to isolate their effects. The top panels include all feedback processes: stellar feedback via supernovae-driven winds, AGN jets and radiative winds, and X-ray heating. Moving downward, we first remove X-ray heating, then AGN feedback, and finally stellar feedback, ending with the no-feedback configuration at the bottom.

X-ray heating preferentially suppresses sSFR in filament-proximal regions by maintaining hot halo gas ($T > 10^{5}$ K) through Compton heating of the intergalactic medium. This process inhibits radiative cooling on scales of $\sim 100$ ckpc, flattening sSFR gradients when absent. A similar trend is seen for SFR, though the effect is more pronounced in sSFR due to its normalisation by stellar mass. Jets, operating on ckpc scales, displace circumgalactic gas through kinetic lobe inflation, creating cavities that reduce gas density in filament-aligned directions. Their removal eliminates the primary quenching mechanism for centrals and cools down the environment of satellites, leading to boosted sSFR near filaments, but the difference between the ``no X'' and ``no AGN'' cases is very subtle considering the uncertainties. Supernovae-driven winds (with velocities $v \sim 100$ km/s) regulate star formation at sub-kpc scales through mass loading factors that eject gas from disk regions. While insufficient to fully counteract AGN effects, stellar feedback preserves residual sSFR excess in satellites when AGN is disabled. This might occur because satellites experience ram pressure stripping in filaments, which concentrates their gas into compact nuclear regions where stellar feedback is not efficient to drive outflows. Altogether, stellar feedback seems to be the key in setting the sSFR excess near filaments, and removing all feedback mechanisms removes all deviations from the median relations. In the no-feedback run, once stellar-mass dependence is removed, star-forming galaxies show no residual modulation of sSFR with distance to filaments. This suggests that, in this regime, star-formation is dominated by the effects of the stellar-mass.

At $z=1$, the trends show an intermediate behaviour between the high-$z$ regime and the low-$z$ regime. Gas accretion is expected to dominate at $z>1$, with filamentary inflows replenishing the interstellar medium of galaxies. Feedback quenching and gas heating become statistically efficient at low $z$ with AGN feedback quenching centrals via momentum-driven cavity formation, and stellar winds maintaining residual star formation in satellites through localised ($<1$ ckpc) gas ejection. We now aim to verify the exact link between baryonic matter accretion and the trends observed in sSFR.

\subsection{Gas accretion}\label{ssec:gas}

\begin{figure}
        \centering
        \includegraphics[width=\columnwidth]{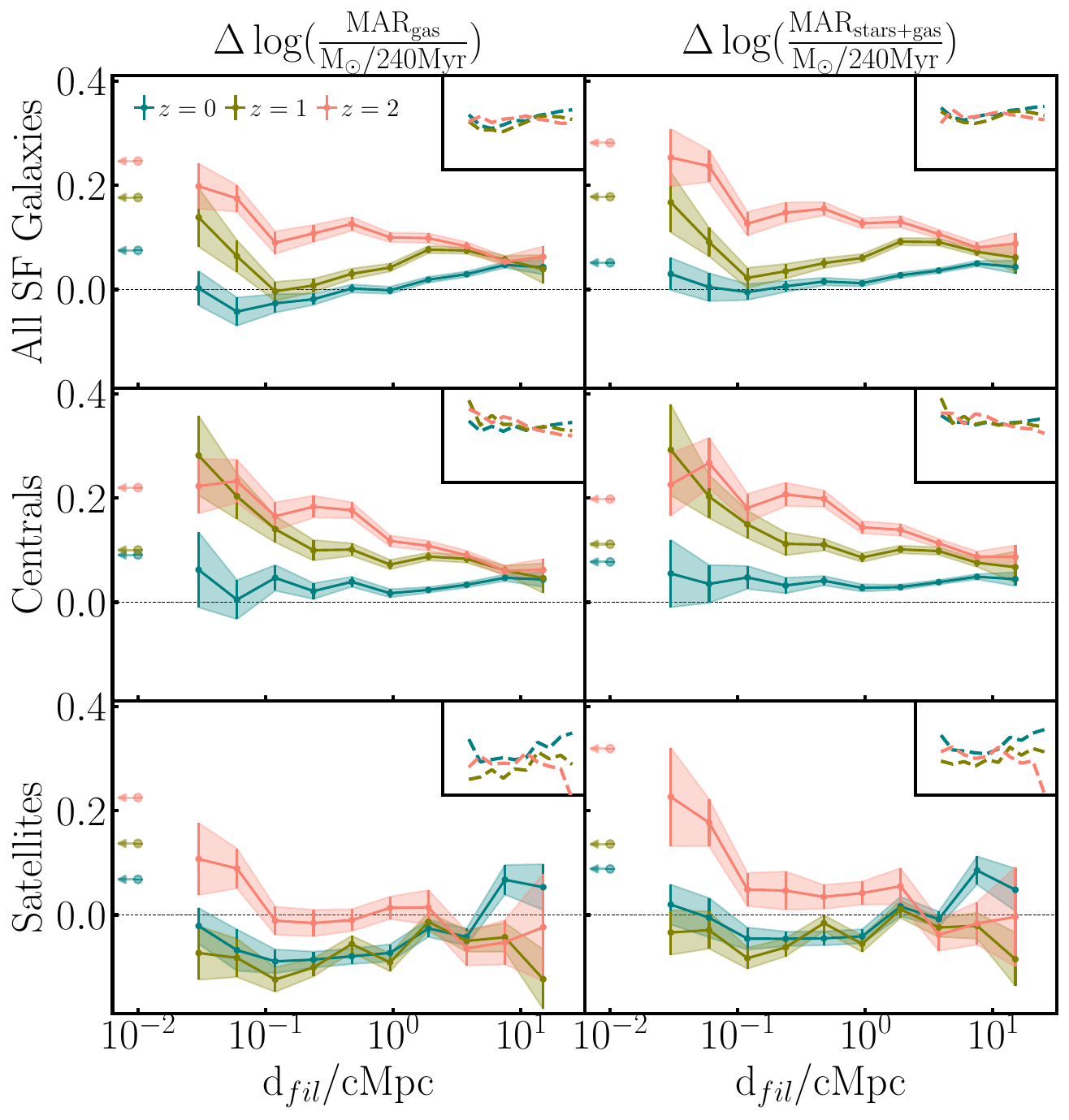}
        \caption{Deviation of the gas mass accretion rate (left panel) and baryonic mass accretion rate (right panel) from the respective best-fit relations in logarithmic scale as a function of the distance to the closest filament for star-forming galaxies in \simba~in bins of distance at $z=0 \text{ (blue), }1 \text{ (green), and }2 \text{ (red)}$ for, from top to bottom, all star-forming galaxies, only central star-forming galaxies, and only satellite star-forming galaxies. The arrows pointing to the left give the value for a bin containing all the galaxies not included in the distance range we study here. For each quantity, the upper right box shows the deviation of the in-bin median from the best-fit relation. Gas and baryonic accretion rates closely track sSFR trends, with a dip at intermediate distances and a recovery in filament cores, especially at $z \leq 1$.}
        \label{fig:MAR}
\end{figure}

The analysis of gas mass and gas and stellar mass accretion rates (MAR), shown in Fig.~\ref{fig:MAR}, provides a direct link between the environmental dependencies of star formation and the physical mechanisms shaping galaxy evolution. To compute these quantities, we track the baryonic particles associated with each galaxy across consecutive snapshots. Gas accretion is defined as the number of gas particles present in a galaxy at a given snapshot that were not part of the galaxy in the previous snapshot. Similarly, stellar accretion is measured by identifying newly formed star particles that were not previously present in the galaxy in any form (neither as stars nor as gas), thus ensuring we capture only genuinely accreted baryonic material. Across all redshifts, the MAR trends and their amplitude variations for star-forming galaxies broadly mirror those of the sSFR and the SFR (see Figs.~\ref{fig:DeltaQ_arranged} and~\ref{fig:distinctionz0}), reinforcing the picture that gas supply regulated by both large-scale structure and internal processes remains the primary driver of star formation activity. A key insight from Fig.~\ref{fig:MAR} is the non-monotonic dependence of MAR on proximity to filaments at $z=1$ and $z=0$. Starting from large distances (several megaparsecs from the filaments) MAR values are relatively high, indicating cold gas supply in low-density environments. As galaxies move inward to intermediate distances ($\sim$0.5–1 cMpc from the filament), MAR steadily decreases. This suppression, most prominent at $z=0$, reflects the growing influence of environmental processes such as shock heating, the development of stable hot gas haloes, and tidal stripping, all of which inhibit fresh gas inflows and contribute to quenching. However, this suppression is not monotonic. As galaxies approach the core regions of filaments, within the innermost $\sim$0.25 cMpc, the MAR increases again. The persistence of the MAR increase close to filaments despite the presence of feedback at low redshift suggests that the densest parts of filaments can still channel significant amounts of cold gas onto galaxies. This inflow, likely tied to coherent large-scale accretion streams along the filament spine, can offset the quenching effects of feedback and maintain star formation even in otherwise hostile environments.

Central galaxies (middle panels of Fig.~\ref{fig:MAR}) display a relatively flat MAR profile near filaments at $z=0$, consistent with the dominant role of AGN jet feedback in suppressing late-time accretion. Yet even for centrals, the MAR upturn at the filament core indicates that the densest environments can still provide a reservoir of cold gas, supporting residual star formation despite strong feedback. At $z=2$, centrals drive the MAR signal, reflecting the efficiency of filamentary accretion during cosmic noon. Satellites, on the other hand, exhibit a much steeper decline in MAR as they approach filaments at $z=0$. Notably, satellites also show a MAR upturn, though less pronounced than for central galaxies, suggesting that even environmentally processed galaxies can access fresh gas if they reside in the very heart of a filament. The transition epoch at $z=1$ marks a shift in the dominant regulatory mechanisms. Centrals still display moderate MAR gradients, indicating that filamentary accretion remains important, while satellites begin to show stronger environmental suppression. The persistence of the central MAR upturn at this epoch highlights the continued relevance of filament cores as preferential sites for gas supply, even as global accretion rates decline and feedback processes intensify. 

The trends identified for all star-forming galaxies qualitatively capture the behaviour of this population regardless of their stellar mass, indicating that the distance-dependent trends found in our study are not driven by the varying mass distribution of galaxies across environments (see Appendix~\ref{appB}). 
As shown in Fig.~\ref{fig:acc_mass}, at $z=1$ and $z=2$, every stellar-mass bin shows the same enhancement of recent accretion at small distances to filaments, driving 
the trend seen for the overall star-forming population. At $z=0$, the low-mass bins continue to follow the characteristic V-shaped dependence on distance: accretion declines from large to intermediate separations and rises again near the filament cores. The trend for the highest stellar mass bin is overall flat, but the uncertainties are large due to small-number statistics. 
The only specific feature appears in the intermediate range $9.5 < \log(M_{\star}/\rm M_{\odot}) < 9.75$, where both the absolute accreted masses and the residuals ($\Delta\log(\rm MAR)$) drop in the innermost distance bins. Galaxies in this mass bin and at very small distances from filaments populate the lower envelope of the accreted-mass–$M_{\star}$ relation and lie close to the transition toward quenched and green-valley populations (see Fig.~\ref{fig:acc_sf_q}) and are also on the low-SFR side of the star-forming main-sequence scatter.
Although these galaxies meet our criteria for star-forming classification (see Sect.~\ref{ssec:quantities}), we speculate that their lower SFR may be physically connected to the suppressed accretion of gas near cosmic web filaments. A more detailed investigation of this specific galaxy population will be addressed in future work.

\section{Discussion}\label{sec:disc}

In the previous section, we have addressed the impact of different galactic and environmental properties on the behaviour of star-forming galaxies as a function of their distance to the closest filament $\mathrm{d}_{fil}$, mostly at $z=0$, 1, and 2. We focused on the effect of recent major mergers, the distinction between central and satellite galaxies, the effect of AGN and stellar feedback, and the recent baryonic mass accretion. We now want to draw a unified scenario that is consistent with the trends seen in our selection of star-forming galaxies, and more specifically the V-shaped sSFR residuals at $z=0$ and the pure large-scale environment influence seen at $z=2$, with $z=1$ being a transition between these two epochs.

\subsection{Interpreting SFR and sSFR trends with distance to the filaments}\label{ssec:intp}

The trends presented in Sect.~\ref{ssec:results} reveal a clear redshift evolution in the relationship between star formation activity and proximity to cosmic filaments. To better understand these patterns, we analyse the behaviour of SFR and sSFR across different redshift intervals. At $z=2$–3, both SFR and sSFR increase toward the filament cores. This behaviour is consistent with efficient gas accretion in dense environments and may reflect the enhanced inflow of cold gas along the cosmic web, as suggested in previous works \citep[e.g.][]{malavasi2022}. Moving to lower redshift ($z=1$), a transition appears, with the sSFR profile forming a shallow minimum near d$_{fil} \sim 0.25$ cMpc. This may indicate the onset of environmental suppression mechanisms acting at intermediate densities. By $z=0$, the sSFR and SFR display a pronounced V-shaped trend, with a decline from low-density regions toward intermediate distances, followed by a recovery near the filament spine. This behaviour suggests that different physical processes operate on different scales. Suppression of star formation at intermediate distances (possibly due to environmental processing), and renewed gas inflows or residual accretion very close to filaments. However, it is important to note that the inner upturn in the sSFR at $z=0$ is a specific feature of the star-forming galaxy population. When quenched galaxies are included in the analysis, this upturn vanishes, and the sSFR decreases monotonically with decreasing distance to the filament. The molecular gas depletion timescale $\tau_{\rm H_{2}}$ exhibits inverse behaviour to SFR at most redshifts. At $z=0$, the increase of $\tau_{\rm H_{2}}$ in the innermost bin, despite rising SFR, may suggest a delayed feedback effect or the onset of a temporary gas replenishment phase. Similarly, the 0.2 dex drop in molecular gas fraction near the filament cores could result from the rapid consumption of gas or environmental stripping processes. To better distinguish between these scenarios and the underlying physical mechanisms, we investigate merger activity, feedback, satellite vs. central contributions, and gas accretion rates in more detail.

The absence of a V-shaped sSFR-$\mathrm{d}_{fil}$ trend among recently merged galaxies at $z=0$ (Fig.~\ref{fig:DeltaQ_mergers}) suggests that major mergers do not drive the behaviour observed in the full star-forming population. Instead, these galaxies show a monotonic decrease in both SFR and sSFR toward filaments, distinct from the population-wide V-shape. This pattern holds at $z=1$ as well, where the sSFR profile remains flat. At $z=2$, a mild increase of sSFR toward filament cores persists, though the amplitude is lower than for the full sample. These results indicate that the redshift-dependent trends in SFR and sSFR with filament proximity are not primarily due to recent major mergers. Although merger rates are higher near filaments at $z \leq 1$, as shown by the increased progenitor counts and major merger fractions in Fig.~\ref{fig:merger_combined}, this does not translate into enhanced sSFR or more prominent environmental modulation. This implies that other mechanisms like variations in gas accretion or feedback effects are likely responsible for the observed environmental trends in star-forming galaxies.

Additionally, the results from Section~\ref{ssec:c_vs_s} show that the dependence of sSFR on the position within the cosmic web at $z=0$ is not uniform across galaxy types. Instead, the signal is strongly driven by environmental substructure and hierarchy. Satellites are the dominant drivers of the observed environmental trends as they exhibit a clear V-shaped sSFR profile with filament distance, particularly when hosted by quenched centrals, highlighting the role of local intra-halo processes or group pre-processing. In contrast, centrals retain star-forming properties largely independent of proximity to filaments. The negligible effect of removing massive halos ($M_h > 10^{13} \rm M_{\odot}$, see Appendix~\ref{appC}) confirms that the signal is not driven by classical cluster environments, but rather by position within the large-scale filamentary network. Finally, we also note that at higher redshifts ($z \gtrsim 1$), the relative importance of satellites decreases due to the lower quenched fractions and satellite abundance, making the environmental modulation of sSFR less pronounced and more reflective of central galaxy behaviour alone.
At $z=0$, for star-forming centrals, the weak V-shaped sSFR-$\mathrm{d}_{fil}$ profile arises naturally from the interplay between feedback and gas accretion. Galaxies in void-like regions evolve largely unaffected, while those at intermediate distances experience a modest decline in sSFR due to feedback-limited gas supply. Closer to filaments, enhanced cold inflows partially offset feedback suppression, leading to the mild upturn observed near the filament cores. This contrasts with satellites of quenched centrals, whose stronger V-shape reflects that satellites embedded in massive hot haloes undergo stronger environmental processing (e.g. gas stripping and heating) at intermediate distances, followed by partial gas re-accretion near filament cores. Their behaviour is therefore consistent with being driven by intra-halo pre-processing rather than large-scale accretion physics.

\subsection{Average environment}\label{ssec:env}

\begin{figure}
        \centering
        \includegraphics[width=\columnwidth]{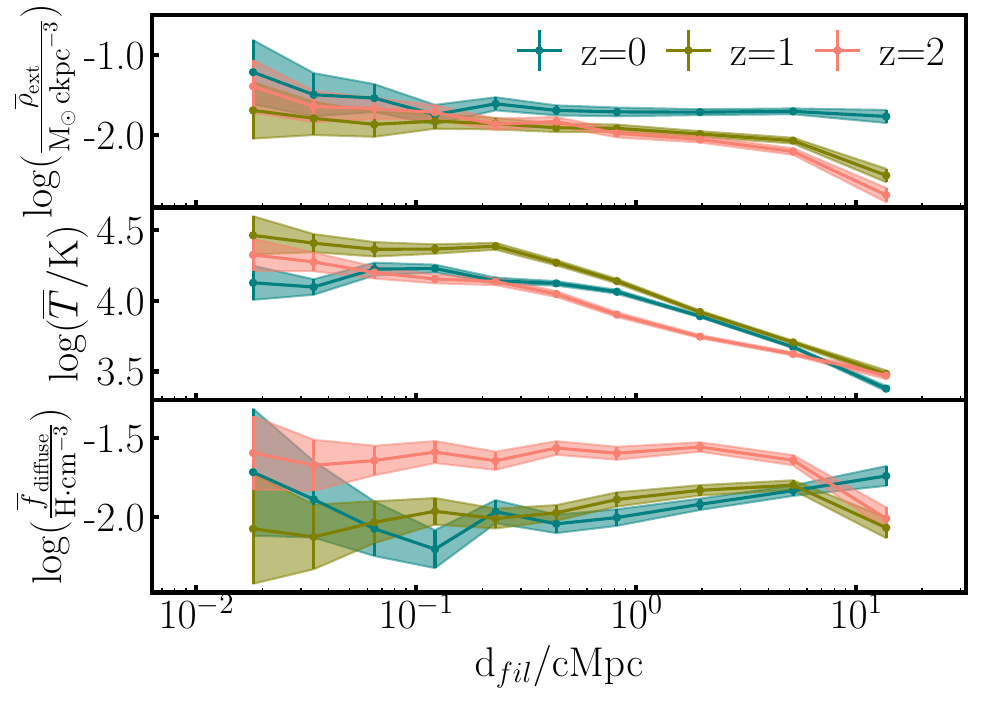}
        \caption{From top to bottom: mean local extragalactic gas density, temperature and diffuse gas fraction at $z=0 \text{ (blue), }1 \text{ (green), and }2 \text{ (red)}$ for star-forming galaxies in \simba~as a function of $\mathrm{d}_{fil}$. At $z \leq 1$, gas density and neutral fraction peak near filaments.}
        \label{fig:rho}
\end{figure}

\begin{figure}
        \centering
        \includegraphics[width=\columnwidth]{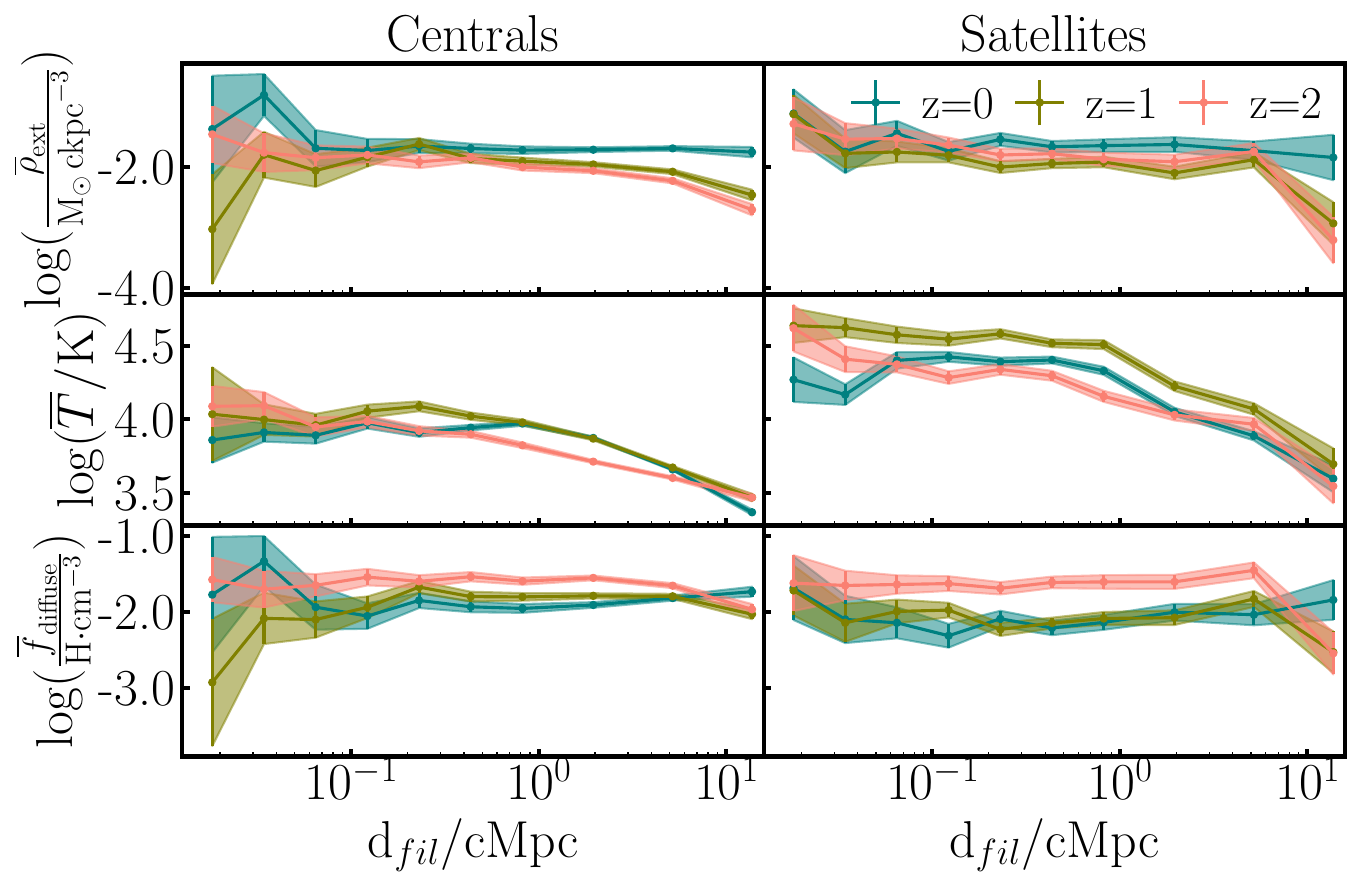}
        \caption{Same as Fig.~\ref{fig:rho} for centrals (left panel) and satellites (right panel).}
        \label{fig:rho_s_c}
\end{figure}

A first approach to assess why the MAR appears to be the main driver of the excess in sSFR and of the trends seen at all redshifts is to look at the local gas environment of galaxies with respect to their distance to the cosmic web. To quantify the local gas environment, we divide the simulation volume into a $256^3$-element grid and compute, in each cell, the average density, temperature, and neutral hydrogen fraction of gas particles not bound to any galaxy as identified by the FoF algorithm. Each galaxy is then assigned the properties of the cell it resides in. The key characteristics of this environment are given in Fig.~\ref{fig:rho}. At all redshifts, the mean gas density increases toward filaments, consistent with filaments acting as primary channels for anisotropic gas inflow from voids toward walls and nodes, thereby sustaining higher ambient densities in these regions. The gas temperature shows redshift-dependent behaviour. At $z=1-2$, it increases toward filaments and then plateaus near their cores, while at $z=0$, it decreases in the innermost bins. This may reflect enhanced radiative losses in denser filamentary gas and a decrease in shock heating efficiency at late times.

At $z=0$, the neutral hydrogen fraction also increases toward filaments and correlates with deviations in sSFR, suggesting a link between diffuse gas availability and ongoing star formation in the local Universe. This correlation is not observed at higher redshifts, where large-scale gas dynamics and accretion dominate over local modulations in the gas environment. As shown in Fig.~\ref{fig:rho_s_c}, central galaxies near filaments generally follow the global density and temperature trends of Fig.~\ref{fig:rho}, but with differences in amplitude and small-scale structure. At $z \geq 1$, they tend to inhabit colder environments, while at $z=0-1$ we find a modest bump–dip feature in density within the innermost $\sim$Mpc that may reflect localised heating or depletion processes. Their temperatures are systematically lower than the global average at high redshift, but converge toward it by $z=0$. For satellites, the density profile remains broadly similar to the overall population but appears flatter at intermediate distances. Their temperature trends are also comparable in shape, with a more pronounced dip–bump sequence emerging at $z=0$, potentially linked to a combination of cooling in pre-infall gas and subsequent heating in group or cluster environments. In diffuse neutral hydrogen fraction, satellites closely follow the global trend at all $z$, whereas centrals remain consistent within $1\sigma$ uncertainties but show a clearer drop at the smallest distances at $z=0$–1. Since our selection minimises intra-halo effects, these trends can be more clearly attributed to the large-scale filamentary context.

A more quantitative assessment of the relationship between local gas density and star formation is provided in Appendix~\ref{appD} where we show that although a weak positive correlation exists between extragalactic gas density and sSFR at all redshifts, this relationship is highly scattered, indicating that variations in local gas density are not sufficient to explain the full environmental modulation traced by cosmic filaments. Filament proximity captures additional aspects of coherent inflows and gas dynamics that are not reflected by static density measurements alone.

\subsection{Star formation in the cosmic web: a multi-scale scenario for centrals and satellites}\label{ssec:scenario}

\begin{figure}
        \centering
        \includegraphics[width=\columnwidth]{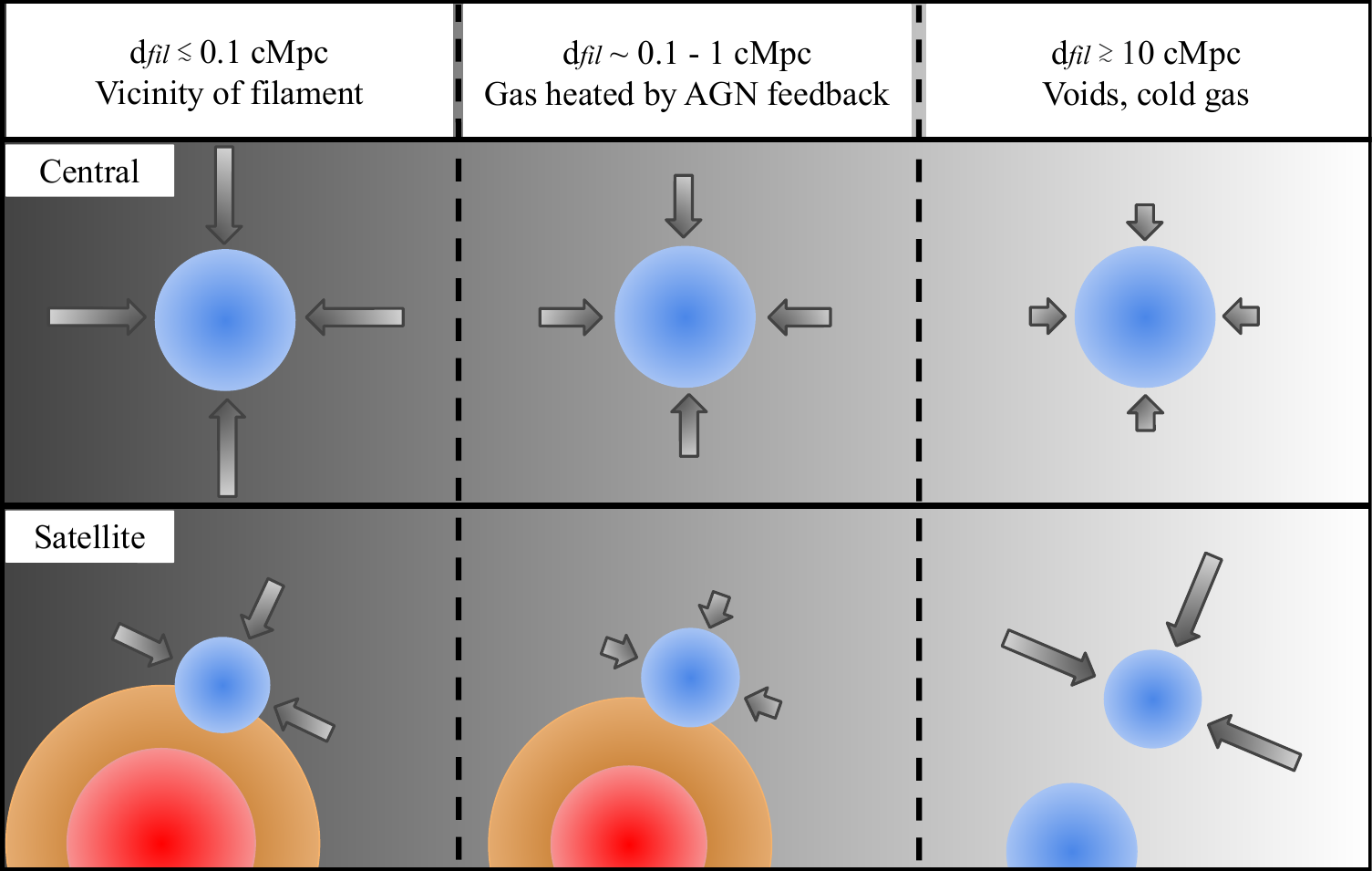}
        \caption{Schematic view of our multi-scale scenarios of gas accretion for central galaxies (upper panel) and satellite galaxies (lower panel) for different typical distances to the filaments from the filaments (left column) to the voids (right columns). The grey arrows give a qualitative representation of the accretion to the star-forming galaxy in blue. For d$_{fil} \lesssim 1$ cMpc, the quenched central galaxy of the satellite is represented in red, with the associated hot CGM in orange. For d$_{fil} \gtrsim 10$ cMpc, most satellites have star-forming central, thus the central galaxy is also in blue. Centrals retain cold gas near filaments, while satellites experience preprocessing and stripping before partial gas re-accretion near filament cores.}
        \label{fig:cartoon}
\end{figure}

Figure~\ref{fig:cartoon} illustrates a multi-scale environmental framework governing star formation for central and satellite galaxies, with centrals and satellites dominating the statistics at $z\sim2$ and $z\sim0$ respectively. For central galaxies, filament proximity enhances cold gas accretion through two mechanisms. Large-scale anisotropic inflows funnel gas from voids into filaments, sustaining elevated molecular hydrogen fractions, and increased local compression at filament interfaces may promote molecular hydrogen formation through cloud collisions and density enhancements. Satellite galaxies, however, exhibit a bimodal response tied to their distance to filaments. Satellites located far from filaments experience moderate gas accretion from the diffuse cosmic web, possibly with lower H$_{2}$ conversion efficiency due to weaker tidal interactions. At intermediate distances ($\sim$0.5–1 cMpc from filaments), satellites of quenched centrals begin to encounter shock-heated gas in group outskirts, which can ionise neutral hydrogen and reduce the supply of cold inflowing gas. This pre-processing, understood as the environmental alteration of satellite properties prior to cluster entry, could explain the observed $\ion{H}{i}$ minima and sSFR suppression in these regions. Finally, within filament-adjacent groups and clusters, satellites can be impacted by additional processes such as ram-pressure stripping and tidal truncation of gas reservoirs. For satellites hosted by quenched centrals, this reduced star formation activity may compete with late-stage accretion, resulting in a complex balance between gas removal and residual star formation activity.

The $z=0$ dominance of the AGN feedback amplifies this dichotomy. Central galaxies sustain star formation through filament-fed accretion until outflows disrupt their CGM, while satellites’ pre-processed gas reservoirs make them vulnerable to rapid stripping upon groups and clusters entry. This can explain the sharp $\ion{H}{i}$–sSFR correlation at $z=0$ as only centrals retain sufficient diffuse gas for feedback-regulated star formation, whereas satellites’ star formation is primarily governed by stripping set by their filament proximity.

Additionally, we explore the dynamical relationship between galaxies and the cosmic web by analysing how short-term movements relative to filaments affect measured environmental trends in Appendix~\ref{appE}. The results demonstrate that the observed dependencies of star formation rates and gas content on filament proximity remain robust even when accounting for galaxies’ motions over a 240 Myr interval at $z=0$, which confirms that the main trends are set by large-scale structure rather than short-term dynamical evolution.

\section{Conclusions}\label{sec:conc}

We have investigated how the star formation activity of galaxies depends on their position within the cosmic web, using the \simba~cosmological simulation from $z=3$ to $z=0$. Using a consistent cosmic web extraction from \disperse~and focusing exclusively on star-forming galaxies, robustly selected via sSFR and gas depletion timescale criteria, we quantified how proximity to cosmic filaments modulates star formation efficiency, gas content, and accretion history across cosmic time. Our main conclusions are as follows:
\begin{itemize}
    \item At high redshift ($z \gtrsim 2$), star-forming galaxies show a significant and smooth dependence on filament proximity suggesting that gas-rich inflows dominate over any environmental suppression.
    \item As redshift decreases, a clear environmental modulation emerges. At $z=0$, we find a characteristic V-shaped trend in sSFR, gas fractions, and depletion timescales as a function of distance to filaments, with minima at intermediate distances ($\sim$0.25–0.5 cMpc) and a recovery closer to the filament spine. This reflects a shift from universal accretion to localised suppression and re-accretion mechanisms.
    \item The observed suppression at intermediate distances is primarily driven by satellite galaxies, which experience environmental preprocessing such as hot halo gas interaction, ram-pressure stripping, and reduced inflow efficiency in the dense outskirts of filaments. Conversely, galaxies at the very cores of filaments can regain gas through coherent cold inflows, partially reviving star formation.
    \item Central galaxies remain largely unaffected by filament proximity at fixed stellar mass, except in the innermost filament regions at $z \lesssim 1$, where residual cold gas inflows may still sustain star formation despite AGN feedback. Satellite galaxies, especially those near quenched centrals, dominate the environmental signal at low redshift.
    \item Major mergers do not drive the environmental trends, as galaxies with recent mergers do not reproduce the V-shaped dependence. Instead, the modulation of gas mass and gas accretion rates closely follows the sSFR trends, indicating that gas availability is the primary regulator of star formation as a function of cosmic web environment.
    \item Analysis of no-feedback runs confirms that AGN and stellar feedback enhance but do not create the environmental trends. The suppression of gas inflows in intermediate-density regions arises primarily from large-scale gas dynamics, with feedback effects amplifying the suppression, especially at low redshift.
\end{itemize}
Together, these results demonstrate that filament proximity alone, even after accounting for galaxy masses and excluding quenched galaxies, imprints a measurable and redshift-dependent modulation on star formation in star-forming galaxies. At low redshift, the maturing cosmic web regulates gas inflows through an interplay of suppression and re-accretion. This leads to a coherent picture in which the large-scale structure not only shapes the assembly of galaxies but also directly regulates their star formation histories across cosmic time.

\begin{acknowledgements}
      The authors thank the referee for insightful comments and suggestions that helped to improve the presentation of the paper. B. J. is supported by a CDSN doctoral studentship through the ENS Paris-Saclay.
      The authors acknowledge the High-Performance Computing Center of the Astronomical Observatory of Strasbourg for supporting this work by providing scientific support and access to computing resources.
      We made extensive use of the {\tt numpy} \citep{oliphant2006guide, van2011numpy}, {\tt scipy} \citep{2020SciPy-NMeth}, {\tt astropy} \citep{1307.6212, 1801.02634}, and {\tt matplotlib} \citep{Hunter:2007} python packages.\\
      The data used and produced in this work are available upon reasonable request to the corresponding author.
\end{acknowledgements}

\bibliographystyle{aa}
\bibliography{main}

\begin{appendix}
\onecolumn

\section{Effect of persistence threshold on galaxy–filament distances}\label{appA}

\begin{figure}[h]
        \centering
        \includegraphics[width=0.7\textwidth]{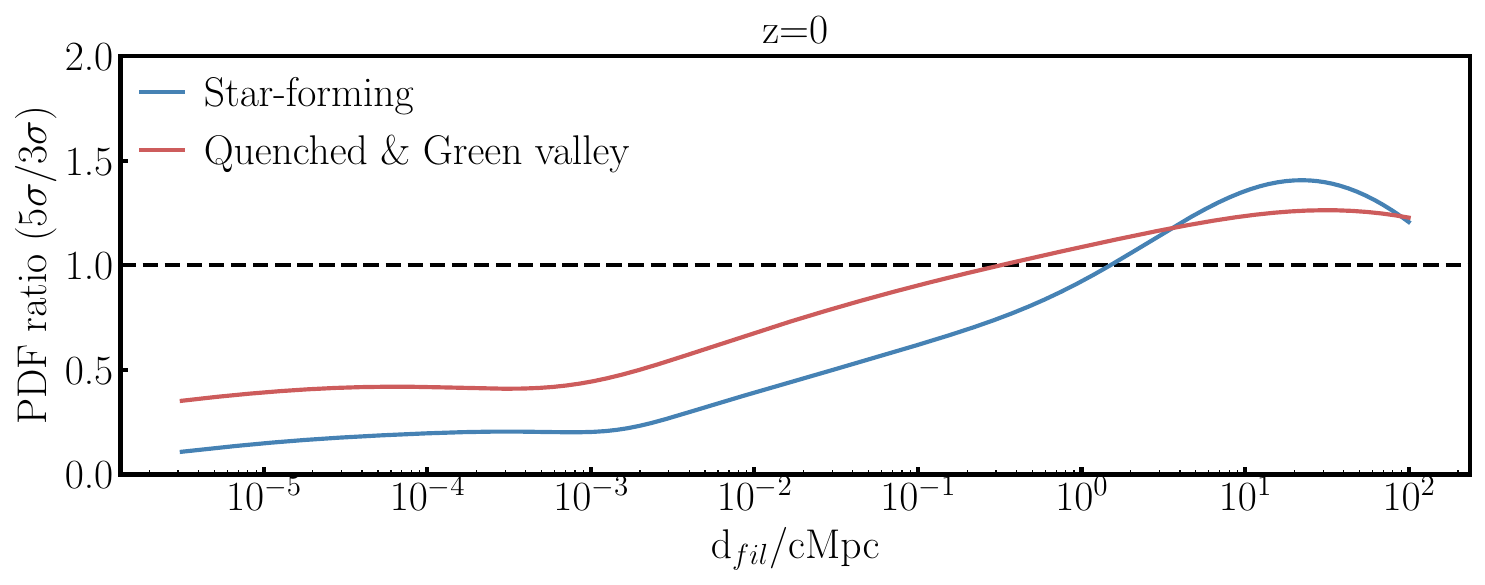}
        \caption{Ratio between the distance distributions obtained with a persistence threshold of $5\sigma$ and $3\sigma$ in \disperse, for quenched galaxies in red and star-forming galaxies in blue. Values below unity indicate a deficit in the $5\sigma$ case, and values above unity an excess.}
        \label{fig:ratio}
\end{figure}

Figure~\ref{fig:ratio} shows the ratio of the distance distributions obtained with a high-persistence ($5\sigma$) and our fiducial ($3\sigma$) \disperse~skeleton, for quenched and star-forming galaxies at $z=0$. A higher persistence threshold preferentially removes the shortest filament segments, reducing the number of galaxies found extremely close to the skeleton. Conversely, the ratio slightly exceeds unity at large separations, indicating a mild excess of galaxies far from filaments in the $5\sigma$ case. The main consequence is not only a change in the number of galaxies at d$_{fil} < 10^{-2}$ cMpc, but also a slight modification of the overall shape of the distance distribution.

As shown in Table~\ref{tab:lowdist}, the number of galaxies at small distances to filaments depends sensitively on the persistence threshold used to build the \disperse~skeleton. For our default value $3\sigma$, only 1133 galaxies lie within this distance range, among which the vast majority (893) are quenched or green valley systems, while only 179 are star-forming. This confirms that galaxies residing extremely close to the skeleton are scarce and are mostly quenched centrals embedded in high-mass halos, consistent with the idea that they trace the nodes of the cosmic web. Indeed, applying a halo mass cut significantly reduces their number as only 708 remain for $M_{h} < 10^{13}\ \rm M_{\odot}$, and just 132 for $M_{h} < 10^{12}\ \rm M_{\odot}$. We verified that increasing the number of smoothing operations in the skeleton construction from one to two removes only 11 galaxies, indicating that the population at d$_{fil}<10^{-2}$ is robust against the level of smoothing. It is much less robust against an increase of the persistence, as increasing it by one approximately cuts this number by a third. The star-forming galaxies that are removed from the d$_{fil}\sim10^{-4}$ cMpc region mostly fall at distances d$_{fil}\gtrsim10^{-2}$ cMpc. Due to their very low number compared to the total number of star-forming galaxies in each bin, they do not significantly impact our analysis. We have verified that the $\Delta\log(\text{sSFR})$ - $\mathrm{d}_{fil}$ relations remain similar to what was obtained in Fig.~\ref{fig:distinctionz0} for both centrals and satellites at $z=0$ with the different configurations.

\begin{table}[h]
    \caption{Number of galaxies with d$_{fil}<10^{-2}$ cMpc at $z=0$ for different persistence thresholds $\sigma$, by types of galaxies and for different halo mass limits.}
    \label{tab:lowdist}
    \centering
    \begin{tabular}{c c c c c c}
        \hline\hline
        Persistence & Galaxies & SF & GV \& Q & Galaxies ($M_{h} < 10^{13}\ \rm M_{\odot}$) & Galaxies ($M_{h} < 10^{12}\ \rm M_{\odot}$)\\ 
        \hline
        $2\sigma$ & 1664 & 340 & 1221 & 1180 & 319 \\  
        $3\sigma$ & 1133 & 179 & 893  & 708  & 132 \\  
        $5\sigma$ & 427  & 34  & 377  & 169  & 11  \\
        \hline
    \end{tabular}
    \tablefoot{Galaxy types are star-forming (SF), including all main sequence and starbursts galaxies, and green-valley and quenched (GV \& Q). The total number of galaxies is given before and after the low-mass cut, and the numbers for each specific types are given as used here, after the low-mass cut. The total number of galaxies is given for all halos, after removing halos with masses above $10^{13}\ \rm M_{\odot}$, and above $10^{12}\ \rm M_{\odot}$.}
\end{table}

\section{Mass accretion}\label{appB}

\begin{figure}[h]
\centering
\includegraphics[width=0.25\textwidth]{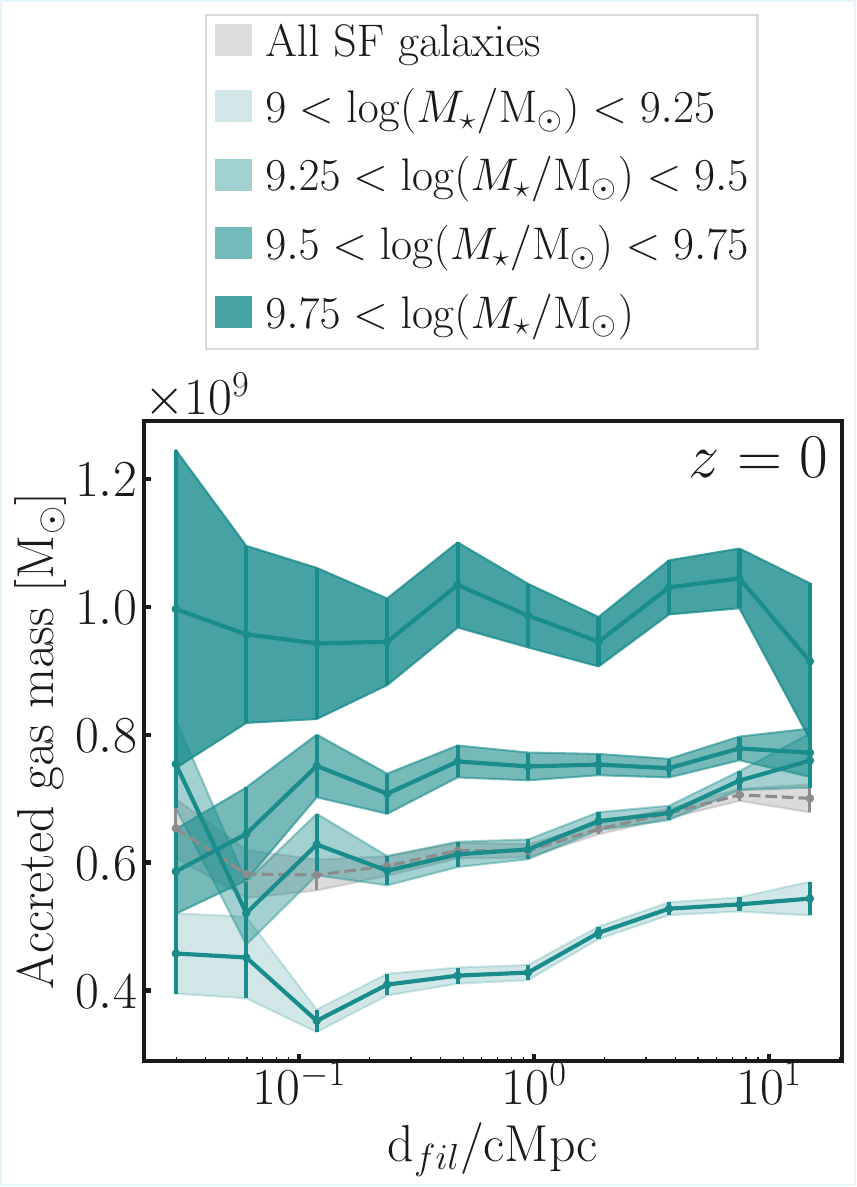}
\includegraphics[width=0.24\textwidth]{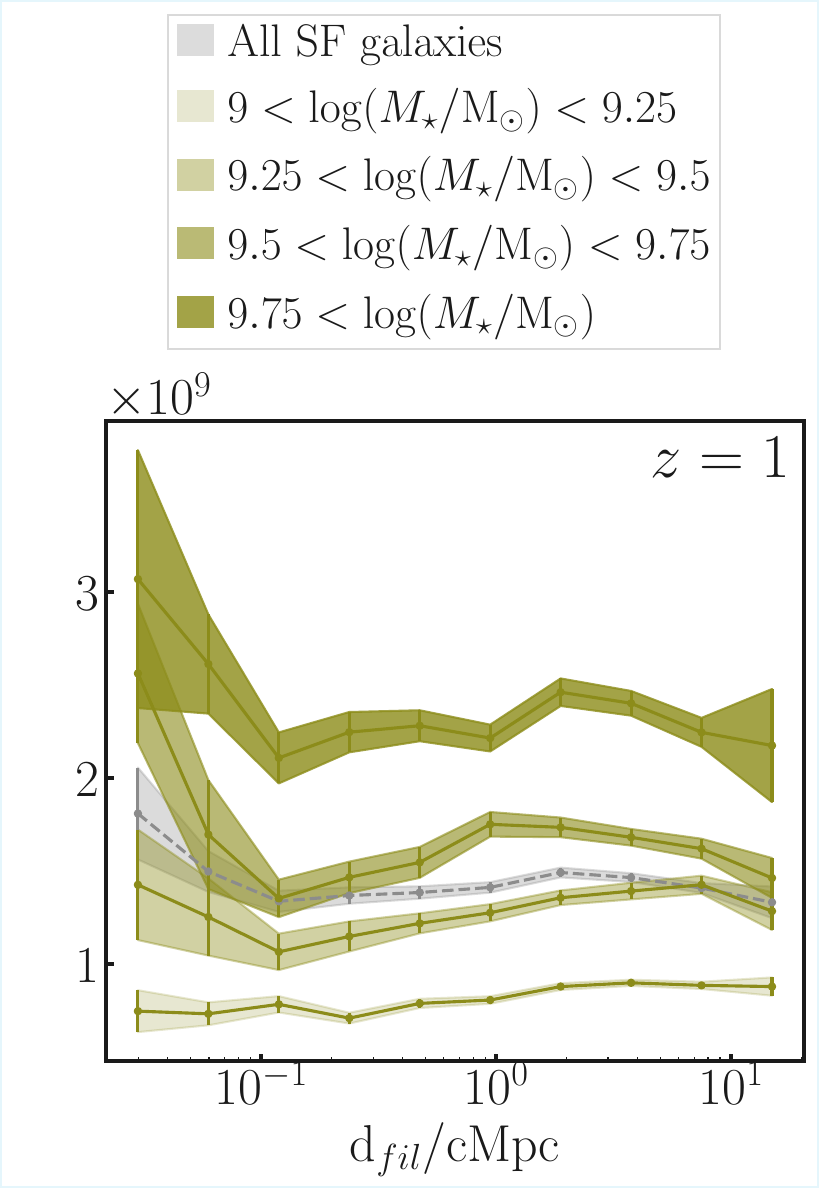}
\includegraphics[width=0.245\textwidth]{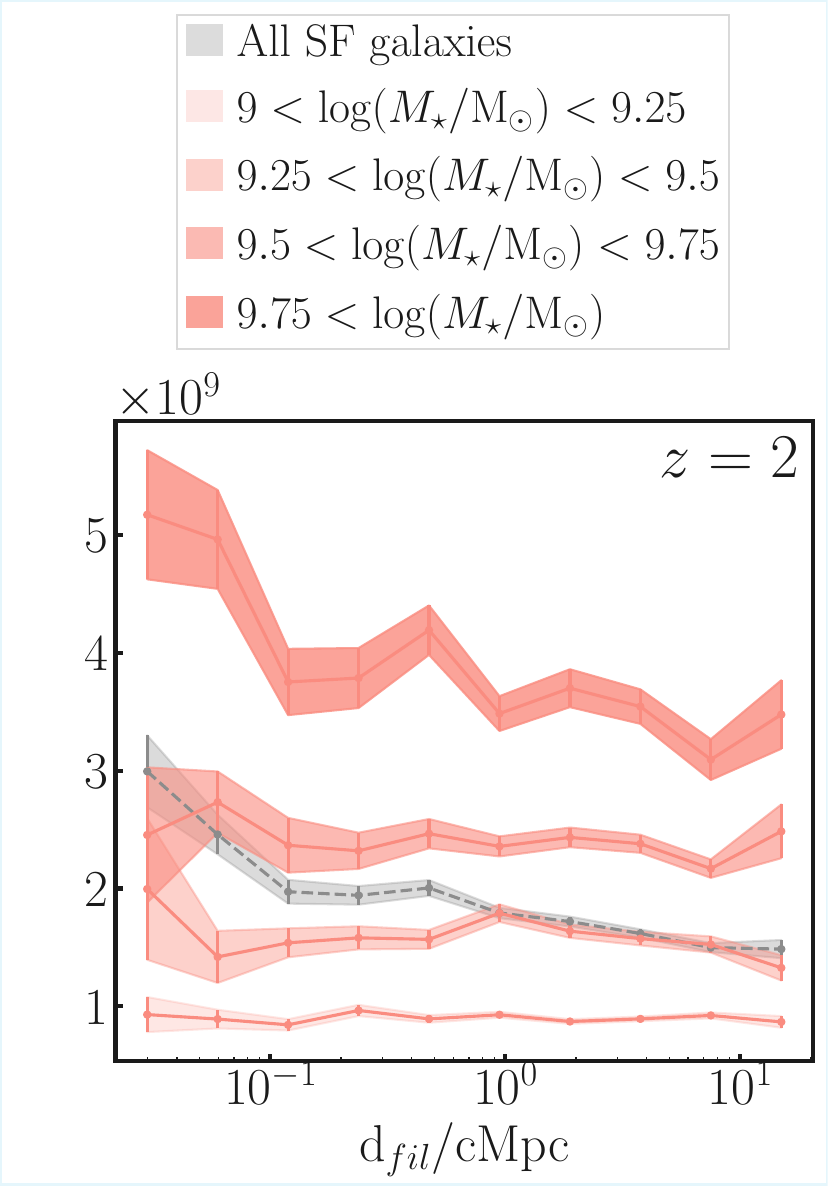}
\includegraphics[width=0.95\textwidth]{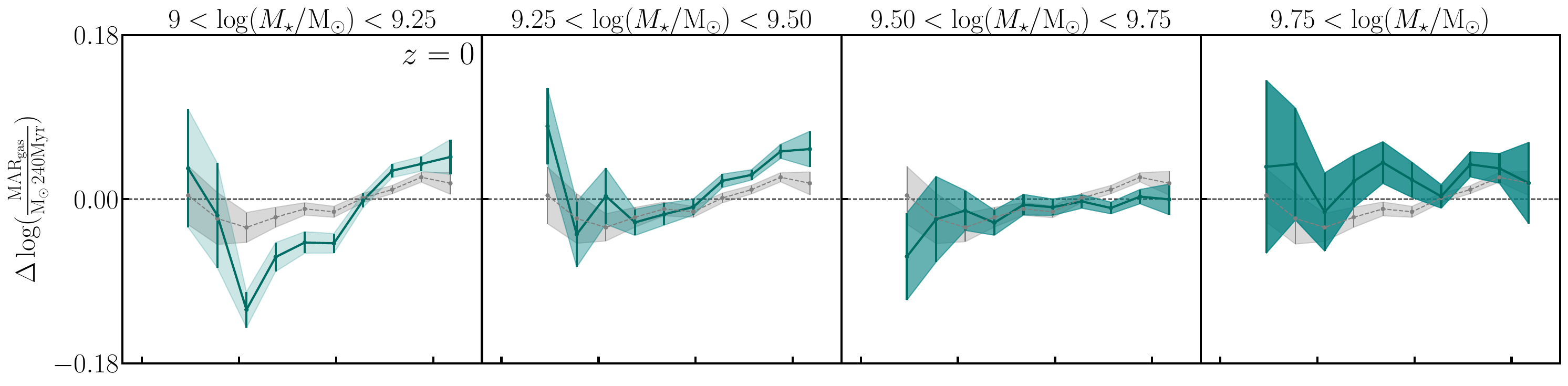}
\includegraphics[width=0.95\textwidth]{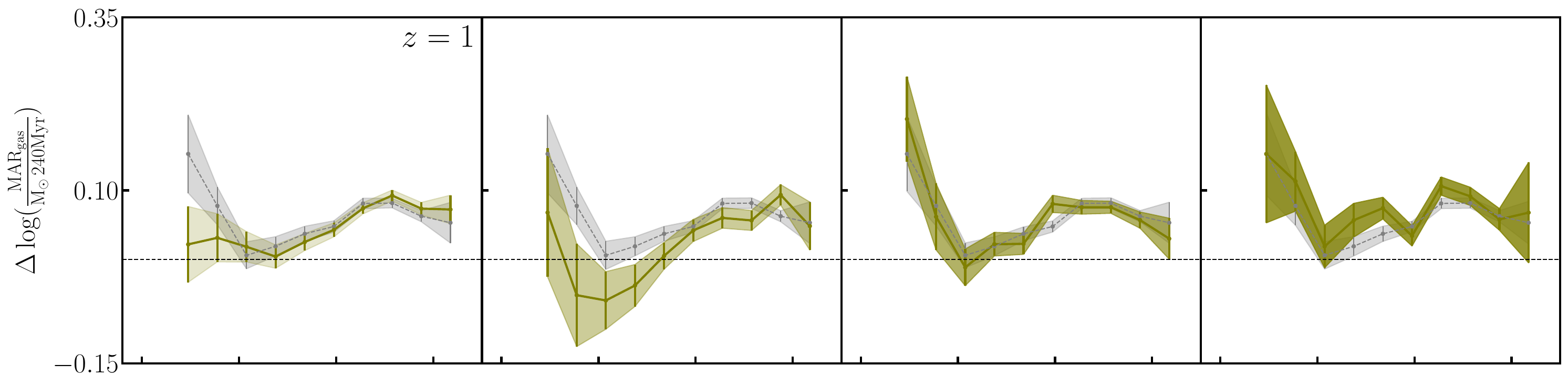}
\includegraphics[width=0.95\textwidth]{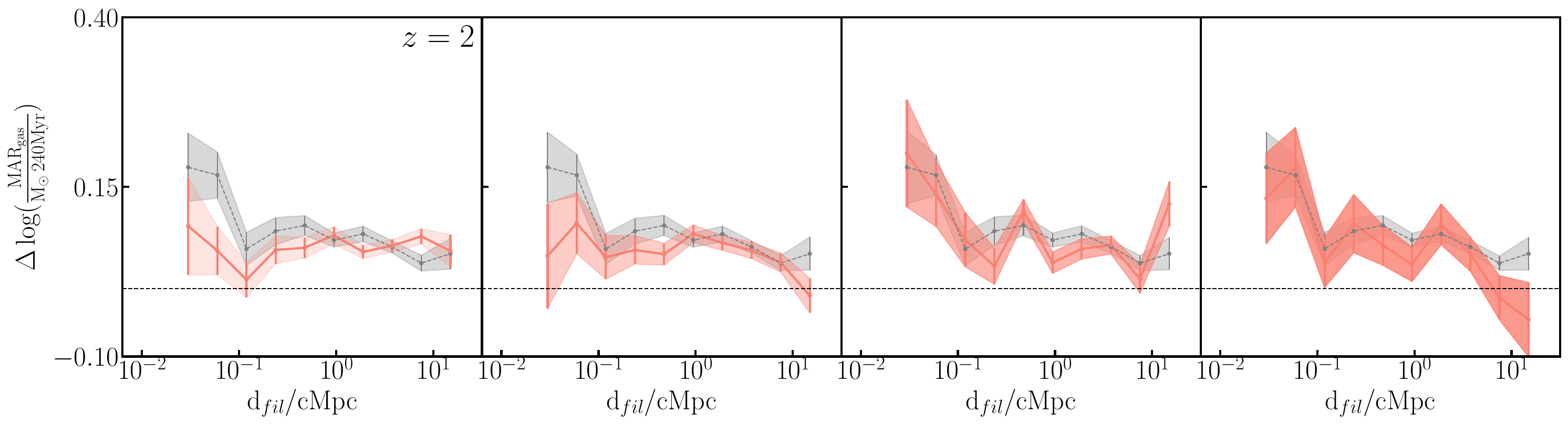}
\caption{The upper panels shows the mean accreted gas mass between two \simba~snapshots at $z=0$ (blue), $z=1$ (green), and $z=2$ (red) as a function of distance to the filaments within different mass bins for star-forming galaxies. The other panels show the residuals in mass accretion rate for galaxies within mass bins, from the lowest masses (left) to the highest masses (right), compared with the residuals for the entire star forming population (in grey, see Sect.~\ref{ssec:gas}) with the same colours for the same redshifts. All bins at all redshifts show trends consistent with the one found for the entire star forming population, but the $9.5 < \log(M_{\star}/\rm M_{\odot}) < 9.75$ bin at $z=0$ is noteworthy as the residuals decrease at low distances, a behaviour not found anywhere else.}
\label{fig:acc_mass}
\end{figure}

\begin{figure}[h]
\centering
\includegraphics[width=0.4\textwidth]{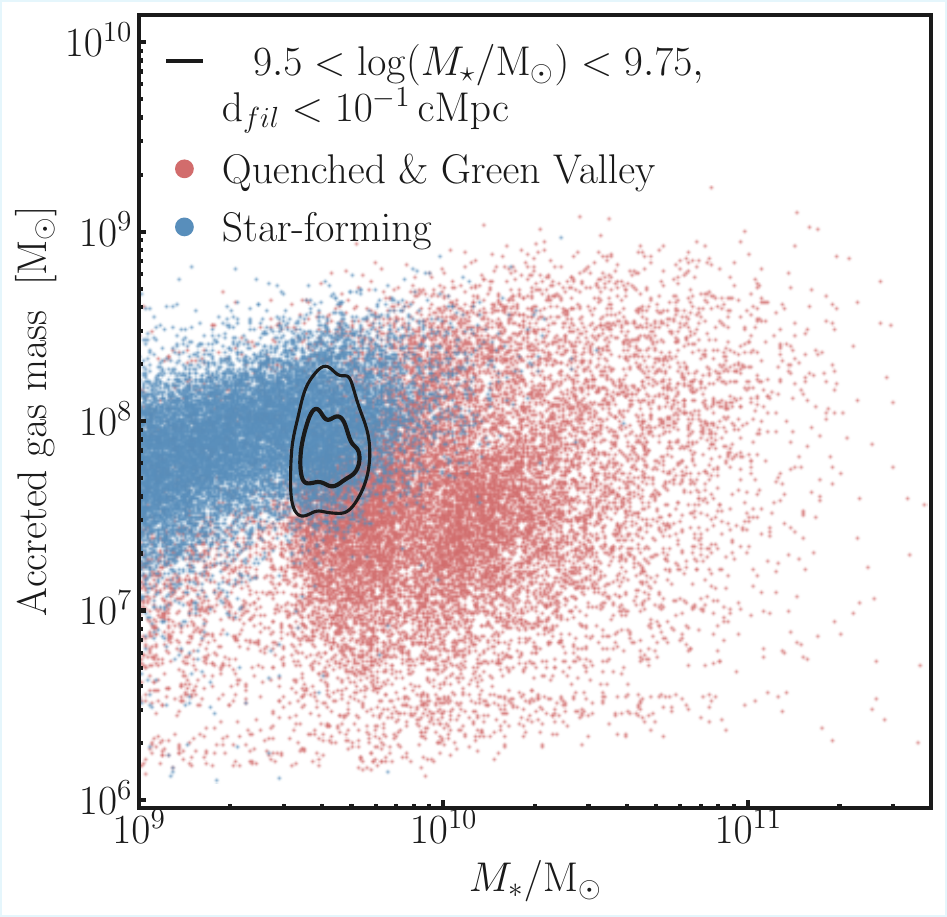}
\caption{Accreted gas mass between two \simba~snapshots at $z=0$ for star-forming galaxies (blue) and quenched and green valley galaxies (red) as a function of stellar mass. The black contours show the $1\sigma$ and $2\sigma$ density distributions for star-forming galaxies with stellar masses between $10^{9.5}\, \rm M_{\odot}$ and $10^{9.75}\, \rm M_{\odot}$ at low distances to the filaments. These galaxies have low accreted gas mass compared with other star-forming galaxies in this mass bin, and follow the general trend towards the quenched population, in agreement to the results at these masses and distances from Fig.~\ref{fig:acc_mass}.}
\label{fig:acc_sf_q}
\end{figure}

To complement the analysis of mass accretion presented in Sect.~\ref{ssec:gas}, we examine here how the dependence of recent gas accretion on the distance to the closest filament varies when splitting the star-forming population into stellar-mass bins. Fig~\ref{fig:acc_mass} summarises the different behaviours at $z = 0$, 1, and 2.

At high redshifts ($z\geq1$), the trends found for the full star-forming population are broadly recovered in every stellar-mass interval. In particular, the residuals with respect to the best-fit accreted–mass-$M_{\star}$ relation tend to rise toward the small separations, consistent with the global enhancement of gas inflow near filament cores discussed in Sect.~\ref{ssec:gas}. This increase is especially clear for the highest-mass bins.

At low redshift ($z = 0$), most stellar-mass bins are in agreement with the expected V-shaped dependence seen for the sSFR and the MAR (see Sects.~\ref{ssec:results} and ~\ref{ssec:gas}), as the mass accretion decreases from large to intermediate distances and rises again close to filaments. The major exception occurs in the intermediate bin $9.5 < \log(M_{\star}/\rm M_{\odot}) < 9.75$, where both the absolute accreted masses and the residuals exhibit a decrease in the three innermost distance bins. This behaviour does not appear in the other stellar-mass bins nor at higher redshifts. Fig.~\ref{fig:acc_sf_q} helps to clarify the origin of this feature. Galaxies in this mass range located at $\mathrm{d}_{fil} < 10^{-1}\, \rm cMpc$ populate the lower envelope of the accreted-gas-mass vs $M_{\star}$ relation and lie much closer to the quenched and green-valley sequence than the rest of the star-forming population. Even at higher masses, the star-forming and quenched populations do not blend in the same fashion, as the high-mass accreting star-forming galaxies are rare but do not show signs of quenching. We note that these galaxies also belong to the lowest-SFR side of the star-forming main-sequence. We speculate that they are evolving toward quiescence and that their suppressed recent gas accretion and low SFR are physically linked, but this remains a plausible explanation rather than a definitive conclusion, since we do not attempt here a full dynamical characterisation of their histories. A separate detailed investigation of this specific population would be necessary to determine whether their suppressed accretion results from environmental processing, internal regulation, or a combination of both.

\section{Effect of large halos}\label{appC}

\begin{figure}[h]
        \centering
        \includegraphics[width=\textwidth]{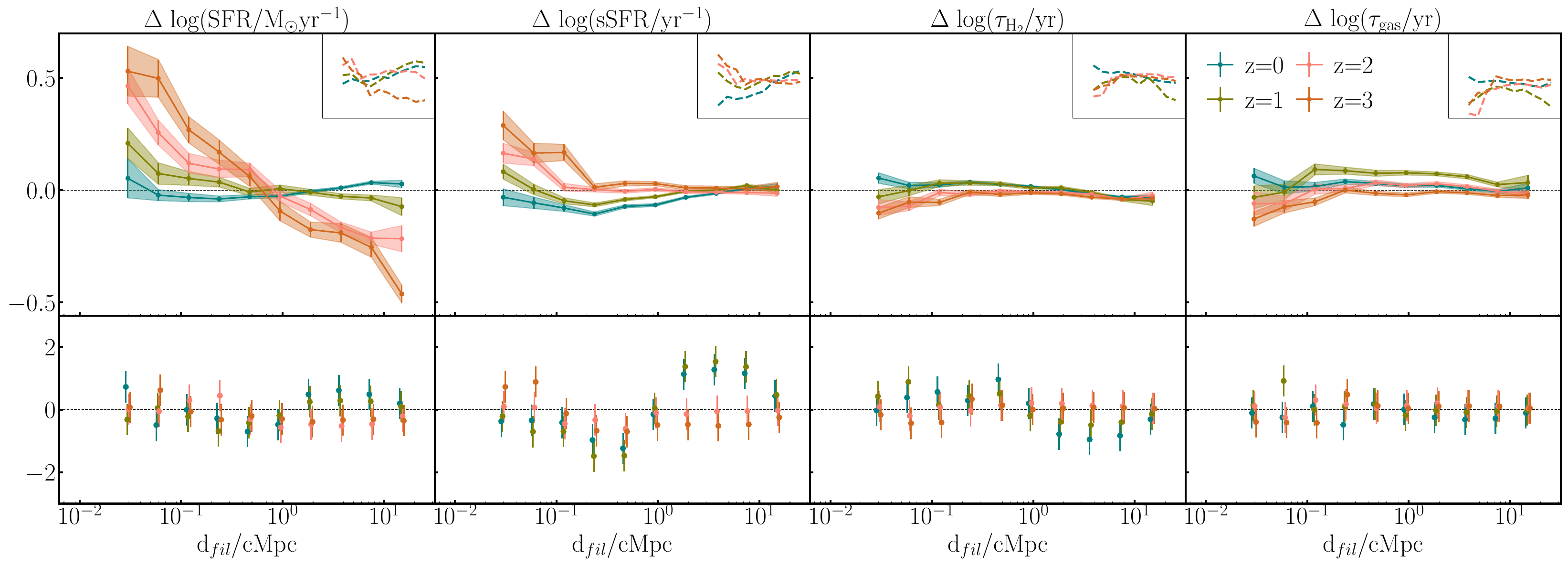}
        \caption{Upper panel: deviation of, from left to right, the mean star-formation rate, specific star-formation rate, molecular hydrogen depletion time, and gas depletion timescale from the respective best-fit relations in logarithmic scale as a function of the distance to the closest filament for star-forming galaxies in \simba~in bins of distance at $z=0$ (blue), 1 (green), 2 (red), and 3 (brown) with the exclusions of galaxies with halo mass M$_{h} > 10^{13}\ \rm M_{\odot}$. The arrows pointing to the left give the value for a bin containing all the galaxies not included in the distance range we study here. For each quantity, the upper right box shows the deviation of the in-bin median from the best-fit relation. Lower panel: standardised deviation of the same quantities between the values obtained in Fig.~\ref{fig:DeltaQ_arranged} and the upper panel. Overall, the results do not deviate significantly from those in Fig~\ref{fig:DeltaQ_arranged}.}
        \label{fig:DeltaQ_nh}
\end{figure}

To verify that the observed environmental trends in galaxy properties are driven by proximity to the cosmic web skeleton rather than residual influence from massive halos, we perform a control analysis by excluding galaxies residing in the most massive halos, i.e. in halos with masses M$_{h} > 10^{13}\ \rm M_{\odot}$. This threshold roughly corresponds to galaxy clusters, where local processes like tidal stripping, ram-pressure stripping, and AGN feedback could dominate over large-scale environmental effects. If the observed correlations between galaxy properties and filament distance were primarily influenced by these dense regions, their removal would substantially alter the measured trends in star-formation rates.  

As shown in Fig.~\ref{fig:DeltaQ_nh}, all key trends remain statistically consistent before and after the exclusion of galaxy clusters from the analysis. The standard deviation between measurements in the original and trimmed samples never exceeds $1\sigma$ in all redshift bins. These results confirm that the geometric imprint of the cosmic web drives the environmental dependencies, unlike localised interactions within massive halos. The persistence of trends suggests that large-scale tidal forces and anisotropic gas flows along filaments regulate galaxy evolution through mechanisms distinct from halo-specific processes.

\section{Is the cosmic web tracing the extragalactic gas density?}\label{appD}

\begin{figure}[h]
        \centering
        \includegraphics[width=\textwidth]{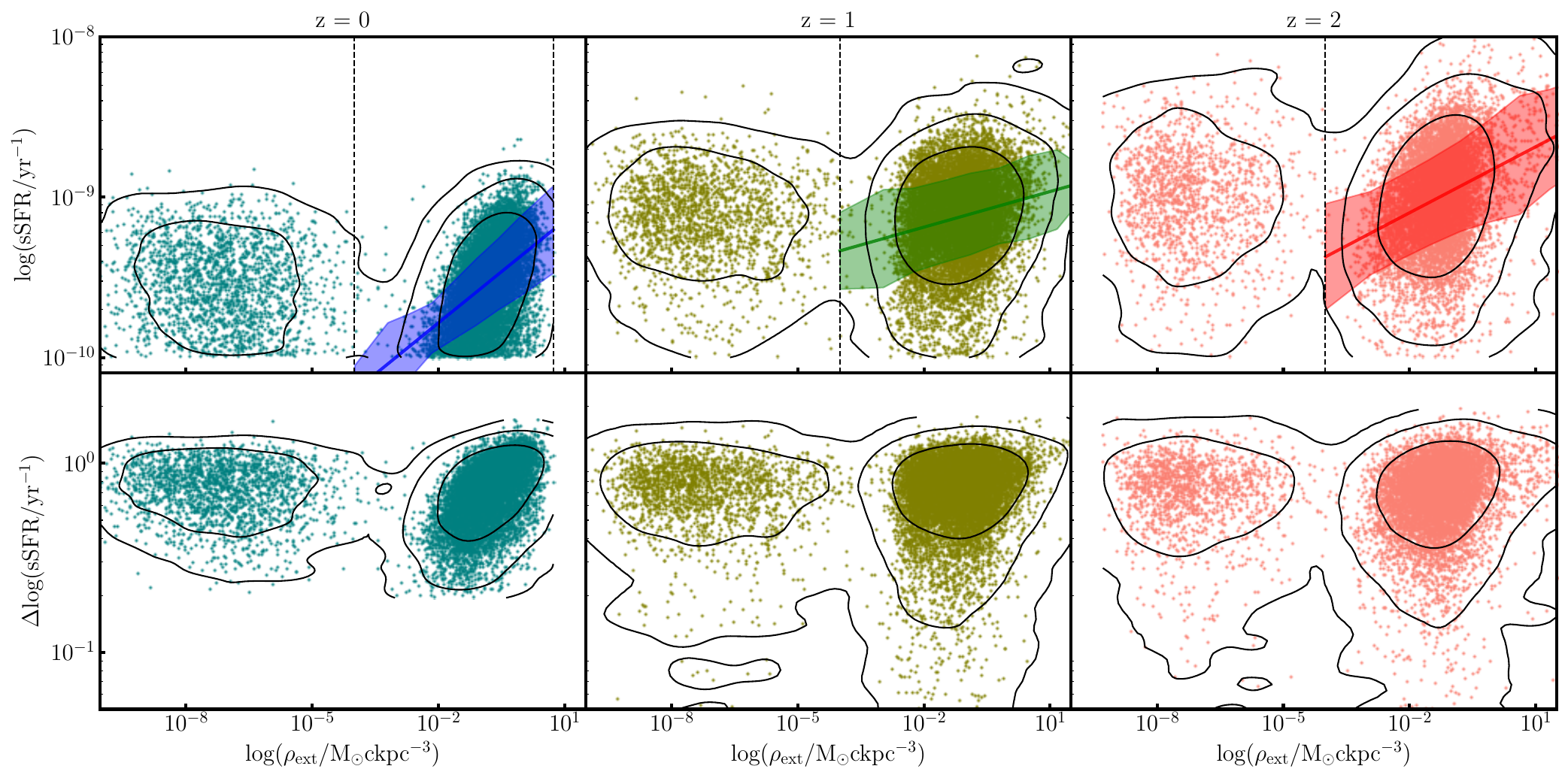}
        \caption{Relation between local extragalactic gas density and star formation activity of star-forming galaxies in \simba~at $z=0$  (blue), 1 (green), and 2 (red). Top row: sSFR as a function of the local external gas density. Bottom row: stellar mass-corrected sSFR offset from the main sequence as a function of the local external gas density. The panels show the 2D scatter, with 1, 2, and 3$\sigma$ density contours in black. We also show the best-fit power-law relation for galaxies with $\rho_{\textrm{ext}}$ > $10^{-4}\ \rm M_{\odot}$ ckpc$^{-3}$), with shaded areas indicating the 1$\sigma$ uncertainty.}
        \label{fig:sSFR_vs_rho}
\end{figure}

To test whether the redshift-dependent modulation of star formation activity with respect to the cosmic web as traced by \disperse~may be partially explained by variations in the surrounding gas environment, we analyse the relationship between the local extragalactic gas density and the star formation efficiency of galaxies. In Fig.~\ref{fig:sSFR_vs_rho}, we show the distribution of star-forming galaxies in \simba~at $z=0$, 1, and 2 in the (log(sSFR/yr$^{-1}$), log$(\rho_{\textrm{ext}}/\rm M_{\odot}$ ckpc$^{-3})$) and ($\Delta$log(sSFR/yr$^{-1}$), log$(\rho_{\textrm{ext}}/\rm M_{\odot}$ ckpc$^{-3})$) planes, where $\rho_{\textrm{ext}}$ is the mean surrounding gas density measured for each galaxy. We overlay the scatter, 1, 2 and 3$\sigma$ contours, and best-fit power-law trends derived with uncertainties.

At all redshifts, we find a broad and highly scattered correlation between gas density and sSFR, with a modest positive slope. Galaxies embedded in denser baryonic environments tend to exhibit higher star formation rates and elevated sSFR relative to the main sequence, but the dynamic range in gas density spans over four orders of magnitude at fixed log(sSFR/yr$^{-1}$), and from one to three for the star formation indicators at fixed density. This indicates the absence of a one-to-one correspondence between gas overdensity and star formation regulation. While local gas density is a necessary condition for sustained star formation, it is not a sufficient driver of the environmental modulation we observe as a function of filament proximity. Proximity to filaments may trace coherent inflows and cold streams not captured by static density metrics, as well as filament-induced compression and turbulence that change the gas characteristics without necessarily increasing the mean gas density. Therefore, although a weak correlation between gas density and sSFR exists in some density ranges, the dominant modulation of galaxy-scale star formation activity by the cosmic web cannot be reduced to a purely local density effect.

\section{Flow of galaxies in the cosmic web}\label{appE}

\begin{figure}
        \centering
        \includegraphics[width=0.49\textwidth]{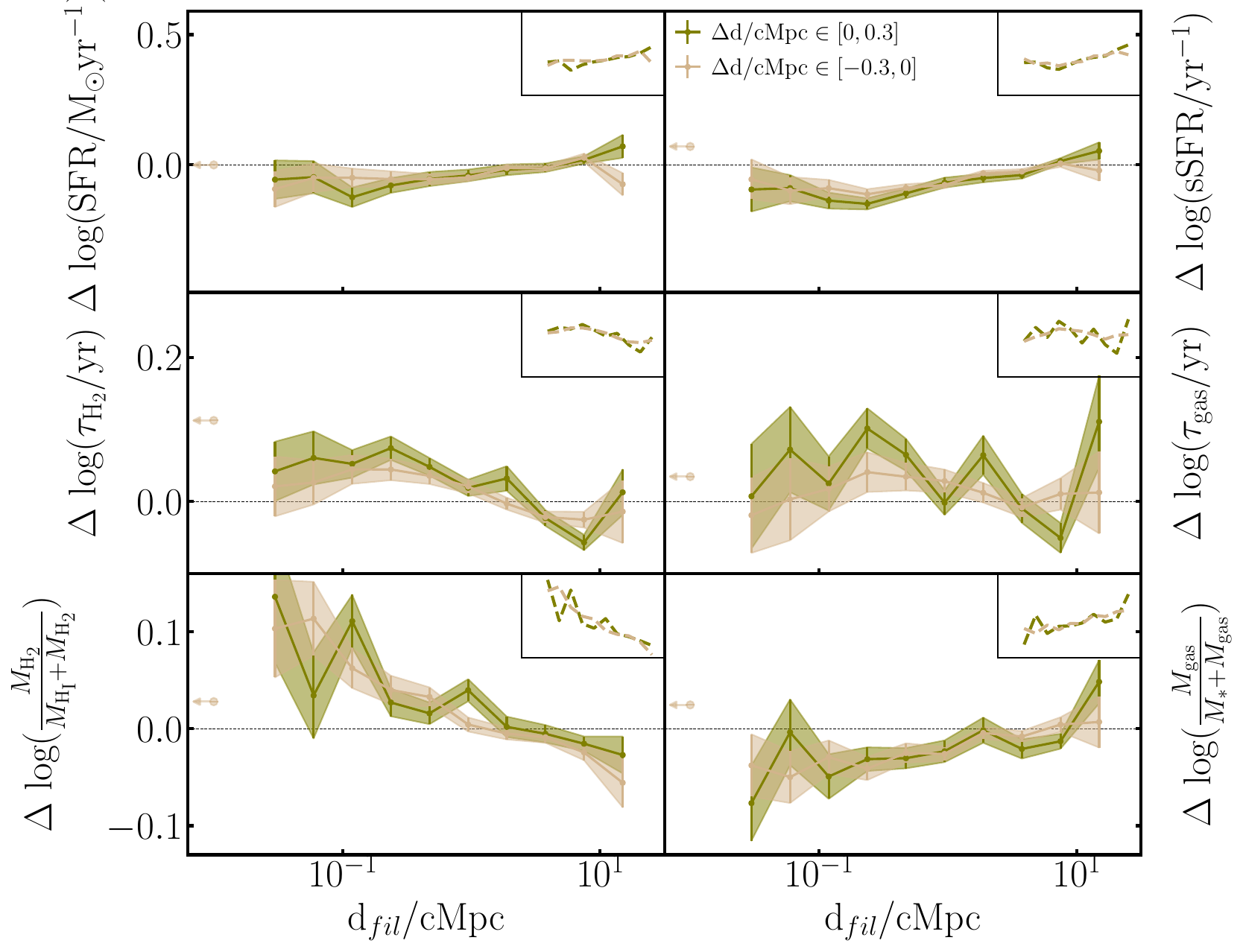}
        \includegraphics[width=0.49\textwidth]{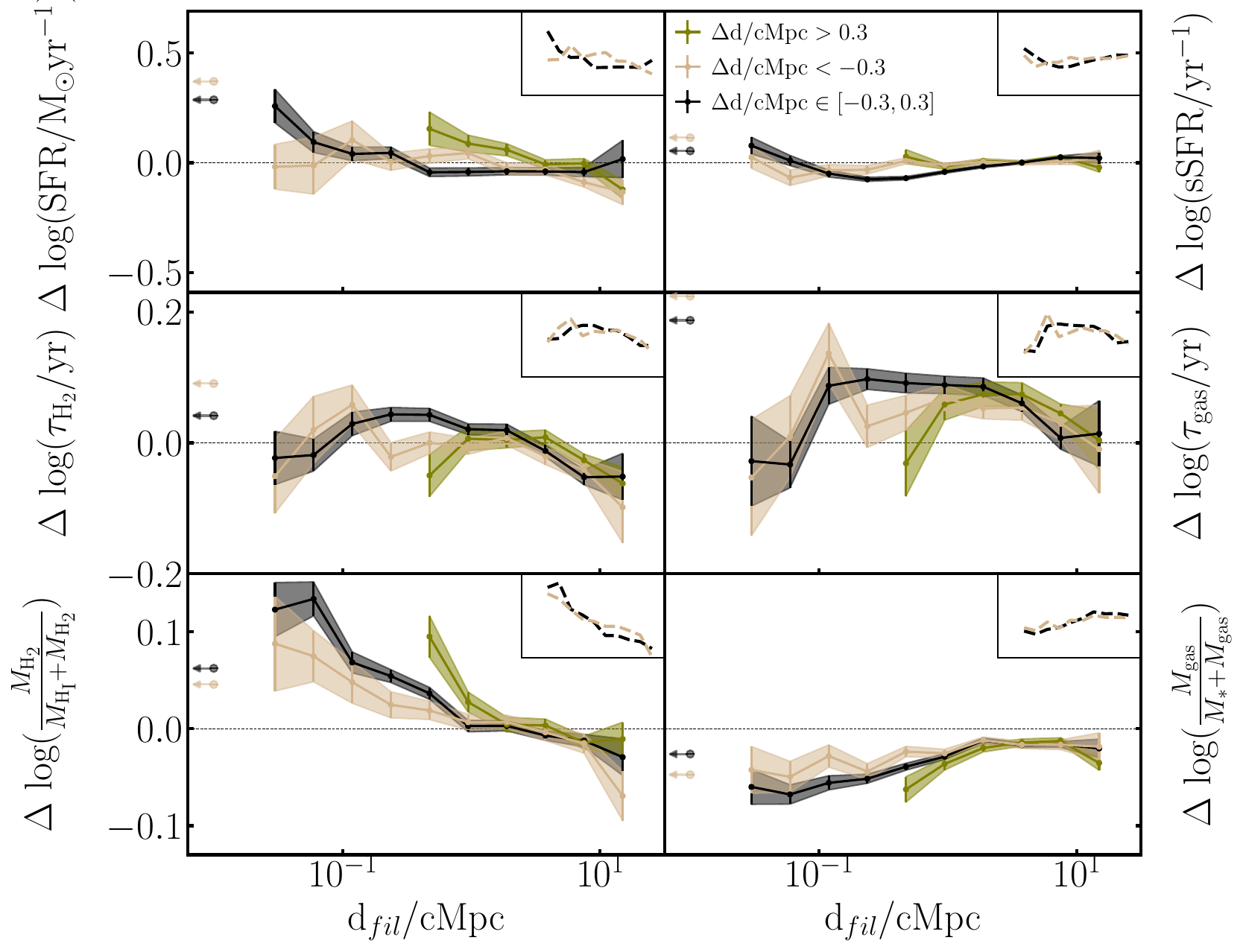}
        \caption{Left panel: same as Fig.~\ref{fig:DeltaQ_arranged} for galaxies which moved further away from the closest filament less than 0.3 cMpc (green) and which got closer to the closest filament less than 0.3 cMpc (brown). Right panel: same as Fig.~\ref{fig:DeltaQ_arranged} for galaxies which moved further away from the closest filament more than 0.3 cMpc (green), galaxies which got closer to the closest filament more than 0.3 cMpc (brown), and galaxies which did not move away or closer to the closest filament more than 0.3 cMpc (black). All the results remain mostly in agreements within the error bars.}
        \label{fig:DeltaQ_movement}
\end{figure}

To better understand the dynamical relationship between galaxies and the cosmic web, we explored how galaxies move relative to their nearest filament over a typical simulation snapshot interval of 240 Myr. This analysis, illustrated in Fig.~\ref{fig:DeltaQ_movement}, allows us to test whether the environmental trends we observe are sensitive to short-term galaxy motions, or if they are instead anchored by the larger-scale structure of the cosmic web itself.

In Fig.~\ref{fig:DeltaQ_movement}, we divide the galaxy sample at $z=0$ according to how much each galaxy’s distance to its closest filament, $\mathrm{d}_{fil}$, has changed over 240 Myr. The distribution of these changes, $\Delta \rm d_{fil}$, is interesting in itself as it is neither purely Gaussian nor sharply peaked, but falls somewhere between a Laplacian and a Lorentzian profile. The full-width at half-maximum of this distribution is about $2 \times 0.3$ cMpc, so we use $\pm 0.3$ cMpc as a natural scale to classify the degree of movement. In the left panel, we focus on galaxies that have moved less than 0.3 cMpc either closer to (brown) or farther from (green) their nearest filament. These represent the majority of the population, which have not experienced significant displacement over the snapshot interval. In the right panel, we highlight galaxies that have moved more than 0.3 cMpc closer to (brown) or farther from (green) their nearest filament, as well as those whose position relative to the filament has remained essentially unchanged (grey).

There is a remarkable consistency between all these sub-samples. Whether a galaxy has moved slightly closer to a filament, drifted a bit farther away, or even shifted by more than 0.3 cMpc, the trends in star-formation rate, gas depletion time, and stellar mass as a function of filament distance are all consistent with each other to within 1$\sigma$. These results are also fully consistent with those presented in Fig.~\ref{fig:DeltaQ_arranged} for the overall population. In other words, the environmental dependencies we measure are robust to the short-term, small-scale motions of galaxies within the web. It is also worth noting that galaxies which move significantly farther from their nearest filament (right panel, green curve) were already, on average, located at larger distances from filaments to begin with. This is entirely consistent with the expectation that galaxies do not rapidly escape the gravitational influence of the large-scale structure; the coherence length of filaments is much larger than the distance a galaxy can travel in 240 Myr. The cosmic web acts as a kind of dynamical scaffold, and while galaxies may experience some local shuffling, their positions relative to filaments remain fairly stable on these timescales.

\end{appendix}

\end{document}